\documentclass[twocolumn,aps,prl,superscriptaddress,tightenlines]{revtex4-1}
\usepackage{amsmath, amsfonts,amssymb}
\usepackage{comment}
\usepackage{soul}
\usepackage{mathtools}			 
\usepackage{physics}
\usepackage{graphicx}
\usepackage{epsfig}
\usepackage{comment}
\usepackage{color}
\usepackage{textcomp}
\usepackage{wasysym}
\usepackage{blindtext}
\usepackage{pifont}
\usepackage[colorlinks,citecolor=blue]{hyperref}
\usepackage{CJKutf8}

\makeatletter
\newcommand*\bigcdot{\mathpalette\bigcdot@{.5}}
\newcommand*\bigcdot@[2]{\mathbin{\vcenter{\hbox{\scalebox{#2}{$\m@th#1\bullet$}}}}}
\makeatother

\begin{document}
	
	\title{Disorder-Induced Strongly Correlated Photons in Waveguide QED}
	\author{Guoqing Tian}
	\affiliation{School of Physics and Institute for Quantum Science and Engineering, Huazhong University of Science and Technology, and Wuhan institute of quantum technology, Wuhan, 430074, P. R. China}
	\author{Li-Li Zheng}\email{zhenglili@jhun.edu.cn}
	\affiliation{School of Artificial Intelligence, Jianghan University, Wuhan 430074, China}
	\author{Zhi-Ming Zhan}
	\affiliation{School of Artificial Intelligence, Jianghan University, Wuhan 430074, China}
	\author{Franco Nori}
	\affiliation{Quantum Computing Center, RIKEN, Wakoshi, Saitama 351-0198, Japan}
	\affiliation{Department of Physics, The University of Michigan, Ann Arbor, Michigan 48109-1040, USA}
	\author{Xin-You L\"{u}}\email{xinyoulu@hust.edu.cn}
	\affiliation{School of Physics and Institute for Quantum Science and Engineering, Huazhong University of Science and Technology, and Wuhan institute of quantum technology, Wuhan, 430074, P. R. China}
	
	\date{\today}
	
	\begin{abstract}
		Strongly correlated photons play a crucial role in modern quantum technologies. Here, we investigate the probability of generating strongly correlated photons in a chain of $N$ qubits coupled to a one-dimensional (1D) waveguide. We found that disorder in the transition frequencies can induce photon antibunching, and especially nearly perfect photon blockade events in the transmission and reflection outputs. As a comparison, in ordered chains, strongly correlated photons cannot be generated in the transmission output, and only weakly antibunched photons are found in the reflection output. The occurrence of nearly perfect photon blockade events stems from the disorder-induced near completely destructive interference of photon scattering paths. Our work highlights the impact of disorder on photon correlation generation and suggests that disorder can enhance the potential for achieving strongly correlated photon.
	\end{abstract}
	\maketitle

	Strongly correlated photons are of importance in a wide range of quantum optical applications, including quantum communication\,\cite{kimble2008quantum,lu2019chip,fn8}, quantum computation\,\cite{knill2001scheme,o2007optical,kok2007linear}, and the study of fundamental quantum mechanics\,\cite{RevModPhys.85.299,RevModPhys.89.021001}. Various methods have been developed to produce such strongly correlated photons in the fields of cavity QED\,\cite{birnbaum2005photon,walther2006cavity,fn2,RevModPhys.87.1379} and waveguide QED\,\cite{chang_colloquium_2018,reitz_cooperative_2022,sheremet_waveguide_2023}. In particular, the 1D waveguide QED platform offers a highly controlled environment for interacting photons with quantum emitters, providing an excellent means to realize strongly correlated photons\,\cite{PhysRevLett.104.203603,versteegh2014observation,PhysRevLett.113.213601,sipahigil2016integrated,brehm2021waveguide}. By leveraging the properties of waveguide structures and the coupling with quantum emitters, the scattered light can exhibit either temporal photon attraction (bunching) or repulsion (antibunching), owing to the interference effects and the intrinsic nonlinearity of quantum emitters, making the waveguide QED platform stand out as a powerful tool for generating and manipulating strongly correlated photons for quantum technologies\,\cite{fn10,fn1,fn11,shitao,prasad_correlating_2020,le2022dynamical,lzg_A,PhysRevLett.132.163602}.

	In practical experimental settings, disorder arising from fabrication limitations is inherently unavoidable. Research on the impact of disorder in quantum systems is an active area of study, e.g., \,\cite{abrahams_scaling_1979,evers_anderson_2008,abanin_colloquium_2019}. Over the past few decades, the effects of disorder have also been extensively studied in the fields of cavity and waveguide QED\,\cite{akkermans_photon_2008,jen_disorder-assisted_2020,fayard_many-body_2021,sommer_molecular_2021,fedorovich_chirality-driven_2022,sauerwein_engineering_2023,viggiano_cooperative_2023,lei_many-body_2023,mattiotti_multifractality_2024,gjonbalaj_modifying_2024}. These works demonstrated that disorder has significant impact on localization-delocalization and the quantum dynamics of the atomic excitations. However, the correlation of scattered photons in the presence of disorder remains relatively unexplored. Crucially, given that the correlation of scattered photons significantly depends on the nonlinearity, namely the level structure of the quantum emitters, this raises an intriguing question: Can the system produce strongly correlated photons, if the disorder in the level structure of quantum emitters is taken into account?
	
	In this work, we address this question and give a positive answer by quantitatively studying the possibility of strong correlation events of scattered photons in a chain of $N$ qubits coupled to a 1D waveguide. The strong correlation events studied in our work are photon antibunching (PA), perfect photon blockade (PPB) and nearly perfect photon blockade (NPPB) events, which have been extensively investigated across various physical systems\,\cite{pb1,sing_photon1,pb2,pb3,fn5,sing_photon2,fn7,fn6,lzg_L}. For a resonant, weak classical input light, the transmission output does not generate antibunched photons when $N=1$, while the reflection output always exhibits PPB due to the Pauli blockade. For $N=2$, neither the transmission nor the reflection output produces antibunched photons. The presence of disorder does not change the possibility of these strong correlation events [see Fig.\,\ref{fig1}(b)].
	
	When the chain contains multiple ($N>2$) qubits, antibunched photons do not occur in the transmission output and only weakly antibunched photons occur in the reflection output. Notably, introducing disorder makes these strong correlation events possible. Especially, we show that the probability of NPPB events remains finite, in stark contrast to ordered chains where NPPB events are absent. The occurrence of NPPB events stems from the destructive interference of the photons scattering path induced by the disorder. Furthermore, the probability of these strongly correlated photon events can be effectively increased by appropriately adjusting the system parameters. Specifically, increasing the chain size can enhance the probability of both PA and NPPB events in the transmission output, and the probability of NPPB events in the reflection output increases with disorder.
	
	\begin{figure}
		\centering
		\includegraphics[width=8.5cm]{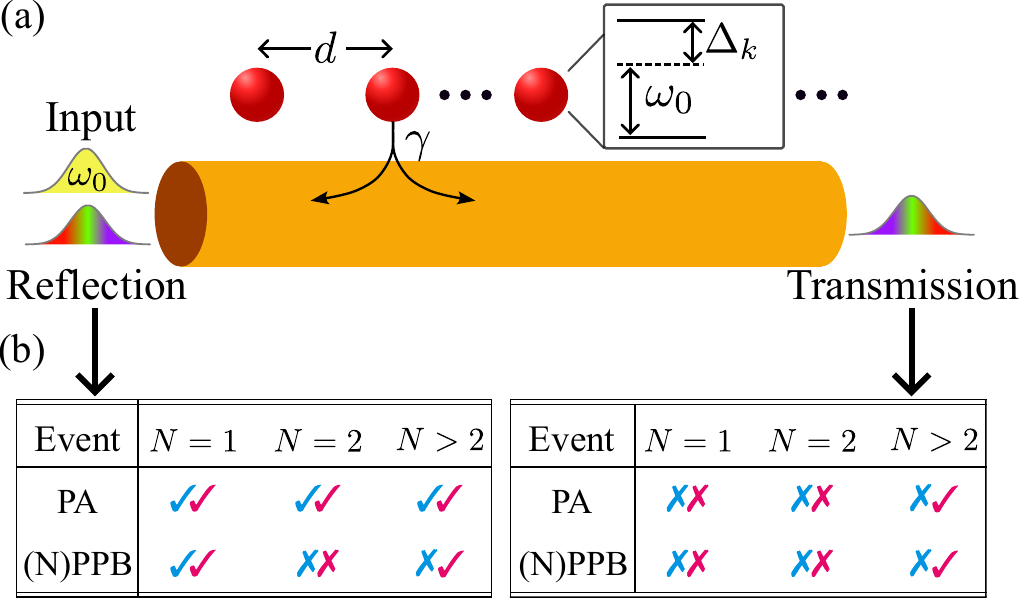}
		\caption{(a) Schematics of a chain of qubits coupled to a 1D waveguide. The qubits have different transition frequencies denoted by $\omega_k=\omega_0+\Delta_k$. The decay rate of each qubit is $\gamma$, and the qubits are uniformly-spaced with distance $d$. For a weak classical input light with frequency $\omega_0$, both the reflection and the transmission outputs can generate strongly correlated photons. (b) PA, PPB, and NPPB events for the reflection and transmission outputs. Here PPB (NPPB) corresponds to $N\le2$ ($N>2$). The corresponding event is possible (denoted by ``\ding{51}") or impossible (``\ding{55}"). Blue (red) color denotes the events for systems without (with) disorder.}\label{fig1}
	\end{figure}

	\emph{Model}.---Let us consider a right propagating coherent pulse with frequency $\omega_0$ (zero bandwidth) and strength $\alpha$ as input light interacting with an chain of $N$ qubits with inhomogeneous transition frequency $\omega_m=\omega_0+\Delta_m$ [see Fig.\,\ref{fig1}(a)]. The detunings of the qubits follow a normal distribution, i.e., $p(\Delta_1,\Delta_2,\cdots,\Delta_N)=\prod_{m=1}^N p(\Delta_m)$, with $p(\Delta_m)=\exp(-\Delta^2_m/2W^2)/\sqrt{2\pi W^2}$. We assume that the qubits only weakly deviate from the resonant frequency $\omega_0$, with the condition $\omega_0/W\gg N$, such that the time evolution of the qubits can be described by the master equation $\dot{\rho}=-i[(H_{\rm eff}+H_d)\rho-\rho (H^{\dagger}_{\rm eff}+H_d)]+\sum_{mn}2(\gamma_{\rm T}\Theta(m-n)+\gamma_{\rm R}\Theta(n-m)+\gamma_{\rm nw}\delta_{m,n})\cos(|m-n|\varphi)\sigma_m\rho\sigma^{\dagger}_n$\,\cite{supp,fn4}.
	Here $H_d=\sum_{m}\sqrt{\gamma_{\rm T}}\alpha(e^{im\varphi}\sigma^{\dagger}_m+h.c.)$, $\gamma_{\rm T}$ ($\gamma_{\rm R}$) is the individual decay rate of each qubit to transmission (reflection) waveguide mode and $\gamma_{\rm nw}$ is the loss to non-waveguide (nw) modes, $\varphi=\omega_0d/c$, with $d$ being the distance between adjacent qubits; $\Theta(x)$ denotes the Heaviside function. The non-Hermitian effective Hamiltonian is (in the rotated frame with respect to $H_0=\sum_{m}\omega_0\sigma^{\dagger}_m\sigma_m$),\nocite{asenjo-garcia_exponential_2017,giant_supp}
	\begin{equation}
		\begin{aligned}\label{eq1}
			H_{\rm eff}&=\sum_{m=1}^{N}\qty(\Delta_m-\frac{i(\gamma_{\rm nw}+\gamma_{\rm T}+\gamma_{\rm R})}{2})\sigma^{\dagger}_m\sigma_m \\ &-i\sum_{m>n}^{N}\qty(\gamma_{\rm T}e^{i|m-n|\varphi}\sigma^{\dagger}_m\sigma_n+\gamma_{\rm R}e^{i|m-n|\varphi}\sigma^{\dagger}_n\sigma_m).
		\end{aligned}
	\end{equation}
	We consider $\gamma_{\rm nw}=0$ and $\gamma_{\rm R}=\gamma_{\rm T}=\gamma/2$ in main text, discussing the impact of losses to non-waveguide modes and the chirality in the Supplemental Materials\,\cite{supp}. We also discuss the effect of finite bandwidth of the input state. We show that the obtained results for a finite-bandwidth input show good agreements with those for zero-bandwidth input, provided that the bandwidth is an order of magnitude below the individual decay rate of qubit\,\cite{supp}. This validates the zero-bandwidth approximation considered in our work; and the requirement of such narrow bandwidth can be experimentally implemented in the state-of-the-art waveguide platforms\,\cite{PhysRevLett.113.213601,kuyken2015octave,bao2024cryogenic}. Hereafter, we choose $\gamma$ as the energy scale and set $\gamma=1$ in our numerical calculations. We only consider $0\le\varphi\le\pi/2$, due to the symmetry of this system.

	The correlation of the emitted field is characterized by the zero-time second-order photon correlation function $g_{\mu}=\langle {a^{\dagger}}^2_{\mu,\rm{out}}(t)a^2_{\mu,\rm{out}}(t)\rangle/\langle a^{\dagger}_{\mu,\rm{out}}(t)a_{\mu,\rm{out}}(t)\rangle^2$, where $a_{\mu,\rm{out}}(t)$, with $\mu=\rm{T}$ ($\rm{R}$) denotes the annihilation operator of the transmission (reflection) mode in the time domain. Utilizing the input-output formalism\,\cite{caneva2015quantum}, the formal expressions of the correlations $g_{\rm \mu}$ in the weak-drive limit, $\alpha\ll1$, can be calculated as\,\cite{supp}
	\begin{align}\label{gt_intef}
		\!\!\!g_{\rm T}\!=\!\frac{|1\!-\!2i\langle {\rm \phi^1_{+}}|\psi^1\rangle\!-\!\langle {\rm \phi^2_{\rm +}}|\psi^2\rangle|^2}{|1-i\langle {\rm \phi^1_{+}}|\psi^1\rangle|^4},\ \ \ g_{\rm R}\!=\!\frac{|\langle {\rm \phi^2_{\rm -}}|\psi^2\rangle|^2}{|\langle {\rm \phi^1_{\rm -}}|\psi^1\rangle|^4}.
	\end{align}
	Here $|{\rm \phi^1_{\pm}}\rangle=\sqrt{1/2}\sum_m \exp(\pm im\varphi)\sigma^{\dagger}_m|G\rangle$ and $|{\rm \phi^2_{\pm}}\rangle=\sum_{m>n} \exp(\pm i(m+n)\varphi)\sigma^{\dagger}_m\sigma^{\dagger}_n|G\rangle$, with $|G\rangle$ being the fully-inverted ground state of qubits; $|\psi^1\rangle=-(H_{\rm eff}^{(1)})^{-1}H_+|G\rangle$ and $|\psi^2\rangle=-(H_{\rm eff}^{(2)})^{-1}H_+|\psi^1\rangle$ are respectively the single- and two-excitation component of the truncated steady-state solution for the qubit ensemble, with $H_+=\sqrt{\gamma/2}\sum_{m} e^{im\varphi}\sigma^{\dagger}_m$. $H_{\rm eff}^{(1)}$ and $H_{\rm eff}^{(2)}$ are the single- and two-excitation sectors of Eq.\,(\ref{eq1}). In the presence of disorder, the probability density functions of $g_{\mu}$ encode the full information of the photon correlations. The definition of the probability density function is given by $P(s)\!=\!\int_{-\infty}^{\infty}\!\!\!\!\!\!\cdots\!\int_{-\infty}^{\infty} \!\!\!\delta\qty(g_{\mu}\!-\!s)p(\Delta_1,\!\cdots\!,\Delta_N){\rm d}\Delta_1\cdots{\rm d}\Delta_N$. The value of $P(s)$ ranges from 0 to infinity, and is proportional to the probability of having a correlation function with value $s$. Especially, $P(\epsilon)$ with $\epsilon\to0$ ($\epsilon=0$) is proportional to the probability of a NPPB (PPB) event. We also define $\mathbb{P}(s<1)\coloneqq \int_0^1 P(s){\rm d}s$, which corresponds to the probability of the field being antibunched.

	\emph{Photon correlations without system disorder}.---Photon correlations in the qubit chains, including the Markov property\,\cite{zheng_L,fang2014,fang2015,shitao,fang2018non,Berman_A} and chirality\,\cite{Mahmoodian} have been extensively studied. For the transmission output, when the qubits are strongly coupled ($\gamma_{\rm nw}=0$) to the waveguide in a non-chiral configuration ($\gamma_{\rm R}=\gamma_{\rm T}$), a straightforward result is that $g_{\rm T}$ is always divergent regardless of the chain size $N$ and distance $d$. The strong photon bunching occurs because resonant qubits block the propagation of single photons from the input\,\cite{atomic_mirror,vds}, hence the output field contains only multiphoton components. While when the qubits are weakly coupled ($\gamma_{\rm nw}\gg \gamma_{\rm R}+\gamma_{\rm T}$) to the waveguide in a perfectly chiral configuration ($\gamma_{\rm R}=0$), the photon statistics of the transmitted light evolving from Poissonian to antibunching and even bunching as the number of qubits increases\,\cite{supp}, which is consistent with the results reported in\,\cite{prasad_correlating_2020}. For the reflection output, $g_{\rm R}=0$ when a single qubit is coupled to the waveguide. This PPB occurs because a single qubit cannot be excited by two photons simultaneously\,\cite{sheremet_waveguide_2023}. When two qubits are coupled to the waveguide, the output light maintains coherent, i.e., $g_{\rm R}=1$. When the chain contains $N>2$ qubits, it is practically impossible to derive a closed-form expression for the correlation function. The numerical result for $g_{\rm R}$ as a function of $N$ and $\varphi$ is given in Fig.\,S4 of the Supplemental Materials\,\cite{supp}. The result shows that, for $N>2$, the reflected photons exhibit weak bunching for $\varphi$ within the range 0 to $0.25\pi$ and weak antibunching for $\varphi$ within the range $0.3\pi$ to $0.5\pi$.
	
	\begin{figure}
		\centering
		\includegraphics[width = 8cm]{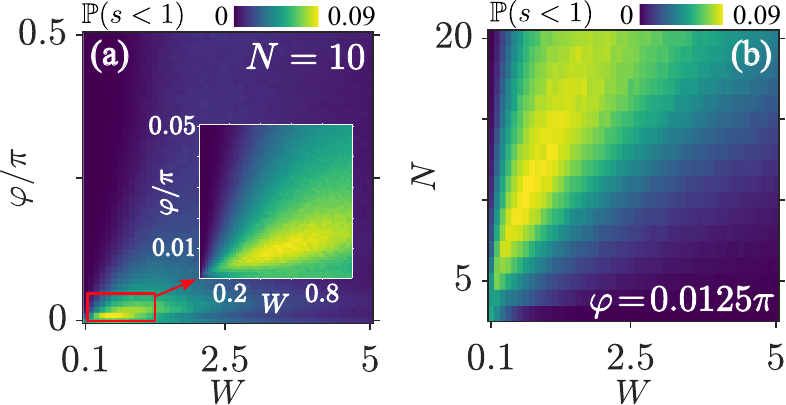}
		\caption{Correlation statistics of the transmission output. Probability of PA versus $\varphi$ and $W$ in (a), and versus $N$ and $W$ in (b). Inset in (a) displays the zoom-in of (a). $N=10$ in (a). In (b), $\varphi=0.0125\pi$, corresponding to the position where $\mathbb{P}(s<1)$ in (a) reach its maximum value. In all plots, the results are obtained from Monte Carlo integration. The numerical details are discussed in the Supplemental Materials\,\cite{supp}.}\label{fig2}
	\end{figure}

	\emph{Transmission photon correlations with system disorder}.---In the presence of disorder, the inhomogeneity in transition frequencies significantly modulates photon correlations in the transmission output. For a chain with a few qubits, we can derive an explicit expression of $g_{\rm T}$. For example, for a single-qubit system the correlation function is given by $g_{\rm T}=(1+4\Delta^2_1)^2/(16\Delta^4_1)$. In contrast to the divergent correlation obtained in a clean chain, $g_{\rm T}$ remains finite provided that the qubit is off-resonant with the input (i.e., $\Delta_1\ne0$). However, one can readily show that $g_{\rm T} > 1$, which indicates that the transmission light is consistently bunched, thereby ruling out the possibility of PA and PPB. In other words, for system with $N=1$ it follows that $\mathbb{P}(s < 1) = P(0) = 0$. For an array with $N=2$, the correlation function is given by
	\begin{align}
		g_{\rm T}=
		\frac{q((\Delta_1+\Delta_2)^2(f_--\cos\varphi)+p)}{(64\Delta^4_1\Delta^4_2(1+(\Delta_1+\Delta_2)^2))},
	\end{align}
	where $q=8\Delta^2_1\Delta^2_2\!+\!f_+\!-\!\cos2\varphi$ and $p=8\Delta^2_1\Delta^2_2(1+(\Delta_1+\Delta_2)^2)+4\Delta_1\Delta_2(\Delta_1+\Delta_2)\sin2\varphi$, with $f_{\pm}=1+2\Delta^2_1+2\Delta^2_2\pm4\Delta_1\Delta_2\cos(2\varphi)-2(\Delta_1+\Delta_2)\sin(2\varphi)$. The correlation function has a minimum value $\min\qty{g_{\rm T}}=1$. Hence, the same conclusion applies to the two-qubit system, i.e., $\mathbb{P}(s < 1) = 0$ and $P(0) = 0$.
	
	\begin{figure}
		\centering
		\includegraphics[width = 8cm]{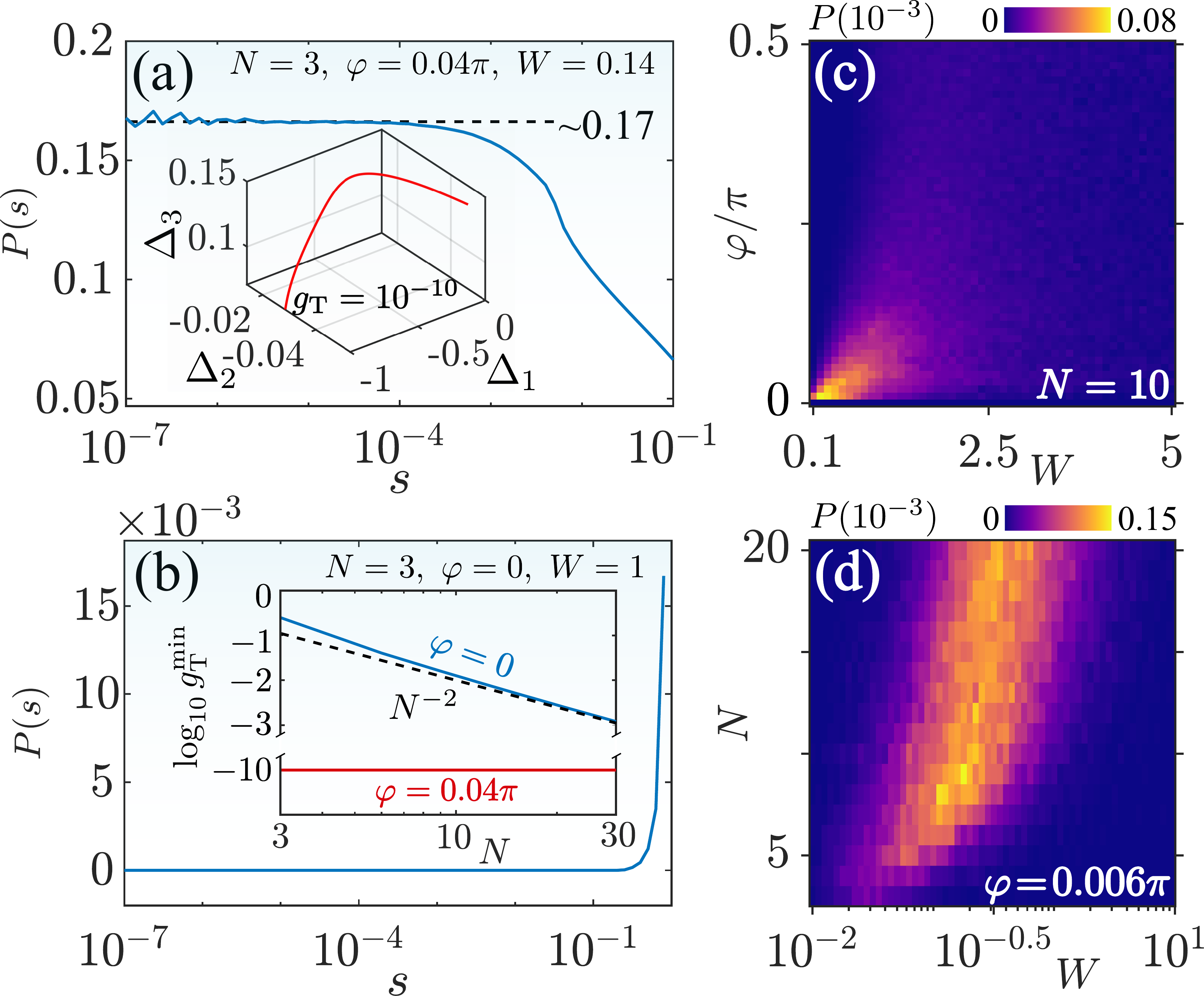}
		\caption{Correlation statistics of the transmission output. (a,b) Probability density functions. The chosen parameters are $\qty{N=3,\varphi=0.04\pi,W=0.14}$ in (a) and $\qty{N=3,\varphi=0,W=1}$ in (b). Inset of (a) shows the solutions $\qty{\Delta_1,\Delta_2,\Delta_3}$ that satisfy $g_{\rm T}=10^{-10}$. The solutions are constrained to $|\Delta_j|\le 1$. Inset of (b) shows $g^{\rm min}_{\rm T}$, corresponding to the minimum value of $g_{\rm T}$, for $\varphi=0$ (blue line) and for $\varphi=0.04\pi$ (orange line) with respect to chain size, dashed line is the numerical fit $N^{-2}$. (c,d) $P(10^{-3})$ versus $\varphi$ and $W$ in (c), and versus $N$ and $W$ in (d). $N=10$ in (c). In (d), $\varphi=0.006\pi$, corresponding to the position where $P(10^{-3})$ in (c) reach its maximum value.}\label{fig3}
	\end{figure}

	When the chain contains more than two qubits, an explicit expression of the correlation function becomes too complex to work out $\mathbb{P}(s<1)$ analytically\,\cite{supp}. In Figs.\,\ref{fig2}(a,b), we present the numerical results of $\mathbb{P}(s < 1)$ under various phases, chain sizes, and disorder strengths. In contrast to the few-qubit array, these results show that antibunched photons can occur in the transmission output. Notably, in the intermediate-disorder regime ($W\sim\order{1}$), $\mathbb{P}(s<1)$ attains its maximum value when the array is highly dense ($\varphi\ll1$) and the chain size is appropriately taken. Such requirement from deeply subwavelength qubit array can be relaxed by the periodicity $\varphi\to\varphi+2\pi$ of the system\,\cite{supp}. Moreover, the deeply subwavelength qubit distance is experimentally feasible in the state-of-the-art superconducting platforms\,\cite{brehm2021waveguide}, so that the condition $\varphi\ll1$ can also be experimentally implemented. For a sparse array, $\mathbb{P}(s<1)$ rapidly decreases with either an increase or a decrease in the disorder strength. In the weak-disorder limit $W\ll 1$, a single photon has a low probability of propagating through the chain, resulting in a low probability of PA. In the strong-disorder limit $W\gg1$, the long-range interaction mediated by photons is largely quenched; consequently, the probability of PA decreases as disorder becomes too strong\,\cite{supp}.

	Regarding the possibility of NPPB events, we first present the probability density function for a system with $N=3$ and $\varphi/\pi=0.04$ in Fig.\,\ref{fig3}(a). Our results indicate that $P(s)$ saturates to a constant value for $s\ll1$, implying that $P(s\to0)\ne0$. This behavior of $P(s)$ is attributed to the existence of specific detuning values $\qty{\Delta_1,\Delta_2,\Delta_3}$ that satisfy the NPPB condition $g_{\rm T}=\varepsilon$ [see the inset of Fig.\,\ref{fig3}(a)], where $\varepsilon$ can be arbitrarily close to 0\,\cite{supp}. According to scattering theory\,\cite{lzg_A}, these NPPB events stem from the nearly completely destructive interference of single-photon scattering paths with probability amplitude $\langle{\rm \phi^1_{+}}|\psi^1\rangle$, the two-photon scattering paths with probability amplitude $\langle{\rm \phi^2_{+}}|\psi^2\rangle$, and the free propagation paths with probability amplitude $1$. These scattering paths are further determined by the transition paths that are governed by the non-Hermitian effective Hamiltonian Eq.\,(\ref{eq1}). A detailed discussion of these destructive effects is provided in\,\cite{supp}. Considering such asymptotic behavior of $P(s)$, hereafter, we adopt $P(10^{-3})$ as the representative value of the probability density function for an NPPB event. In Figs.\,\ref{fig3}(c,d), we present $P(10^{-3})$ for different system parameters. Our results show that NPPB events can occur, provided that $\varphi\ne0$. In the Dicke-limit $\varphi=0$, however, we find $P(10^{-3})=0$ for relatively small chain sizes [see Fig.\,\ref{fig3}(b)]. In this case, $P(10^{-3})$ vanishes because, unlike the systems with $\varphi\ne0$ where $g_{\rm T}$ can be arbitrarily close to 0, the correlation function for $\varphi=0$ can only attain a minimal value that scales as $N^{-2}$ with the chain size [see the inset of Fig.\,\ref{fig3}(b)]. This indicates that $P(\epsilon)=0$ for $\epsilon\lesssim N^{-2}$.
	
	\begin{figure}
		\centering
		\includegraphics[width = 8.5cm]{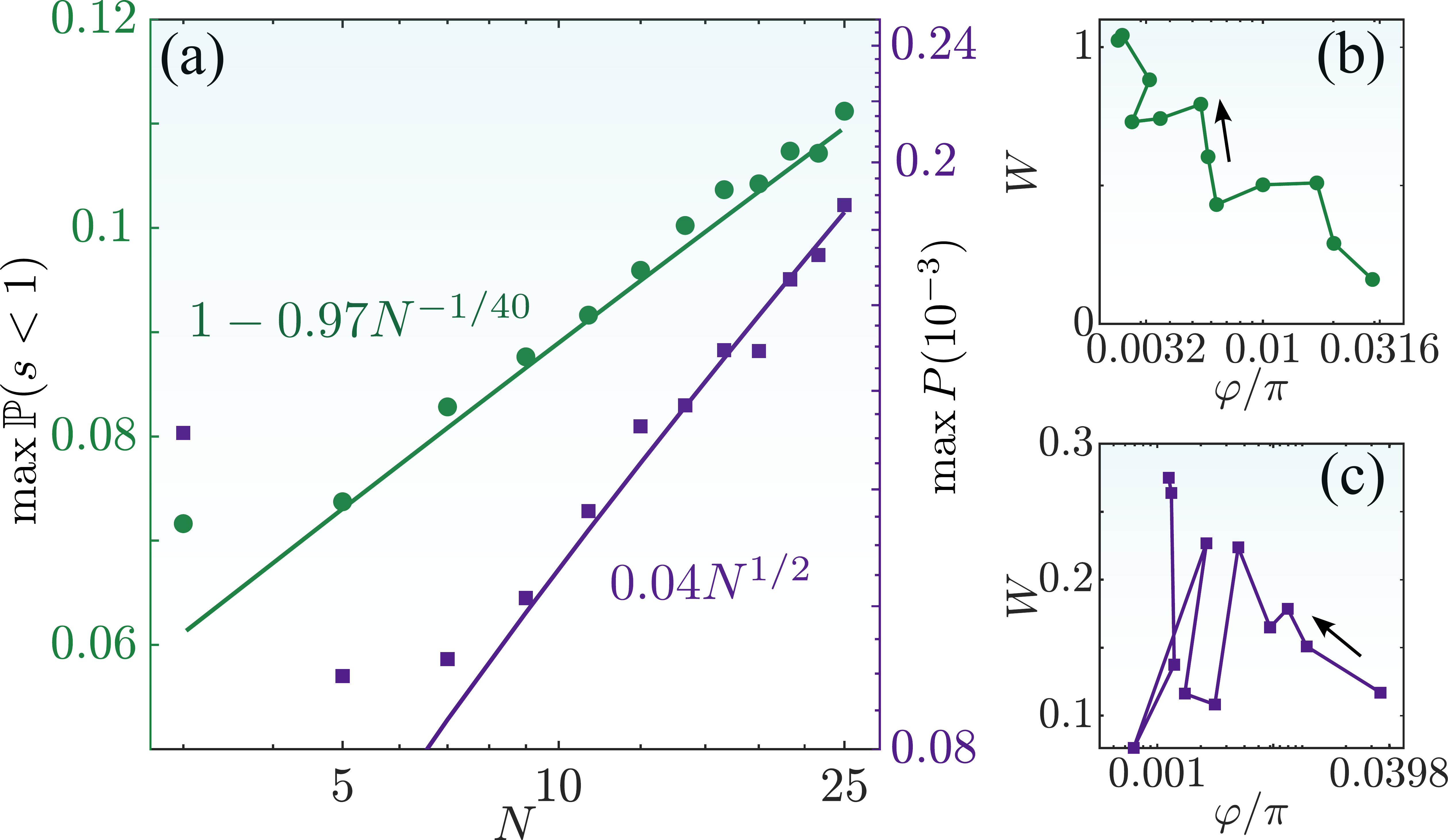}
		\caption{Correlation statistics of the transmission output. (a) The maximum values of $\mathbb{P}(s<1)$ and $P(10^{-3})$ versus $N$. Green circles (purple squares) display the maximum values of $\mathbb{P}(s<1)$ ($P(10^{-3})$) versus $N$. Green and purple solid lines are the numerical fits of $1-0.97N^{-1/40}$ and $0.04N^{1/2}$, respectively. (b) and (c) display $W$ and $\varphi$, respectively, where $\mathbb{P}(s<1)$ and $P(10^{-3})$ reach their maximum values. Black arrows point in the direction of increasing $N$. Inset of (c) displays the zoom-in of (c).}\label{fig4}
	\end{figure}
	
	Our results shown in Figs.\,\ref{fig2}-\ref{fig3} demonstrate that both $\mathbb{P}(s<1)$ and $P(10^{-3})$ can attain their optimal (maximum) values when system parameters are suitably tuned. We subsequently investigate how these optimal values depend on the system parameters. As shown in Fig.\,\ref{fig4}(a), the maximum values of $\mathbb{P}(s<1)$ and $P(10^{-3})$ exhibit power-law scaling with the number $N$ of qubits according to $(1-0.97N^{-1/40})$ and $0.04N^{1/2}$, respectively. In addition to power-law scaling, the dependence of these optimal values on $W$ and $\varphi$ exhibits both similar and distinctive features. Specifically, as the number of qubits increases, the optimal values of $\mathbb{P}(s<1)$ and $P(10^{-3})$ are reached at lower values of $\varphi$; however, achieving the maximum $\mathbb{P}(s<1)$ requires a stronger disorder strength, whereas the maximum $P(10^{-3})$ is obtained when the disorder strength is around 0.15 [see Figs.\,\ref{fig4}(b,c)].
	
	\begin{figure}
		\centering
		\includegraphics[width=8.4cm]{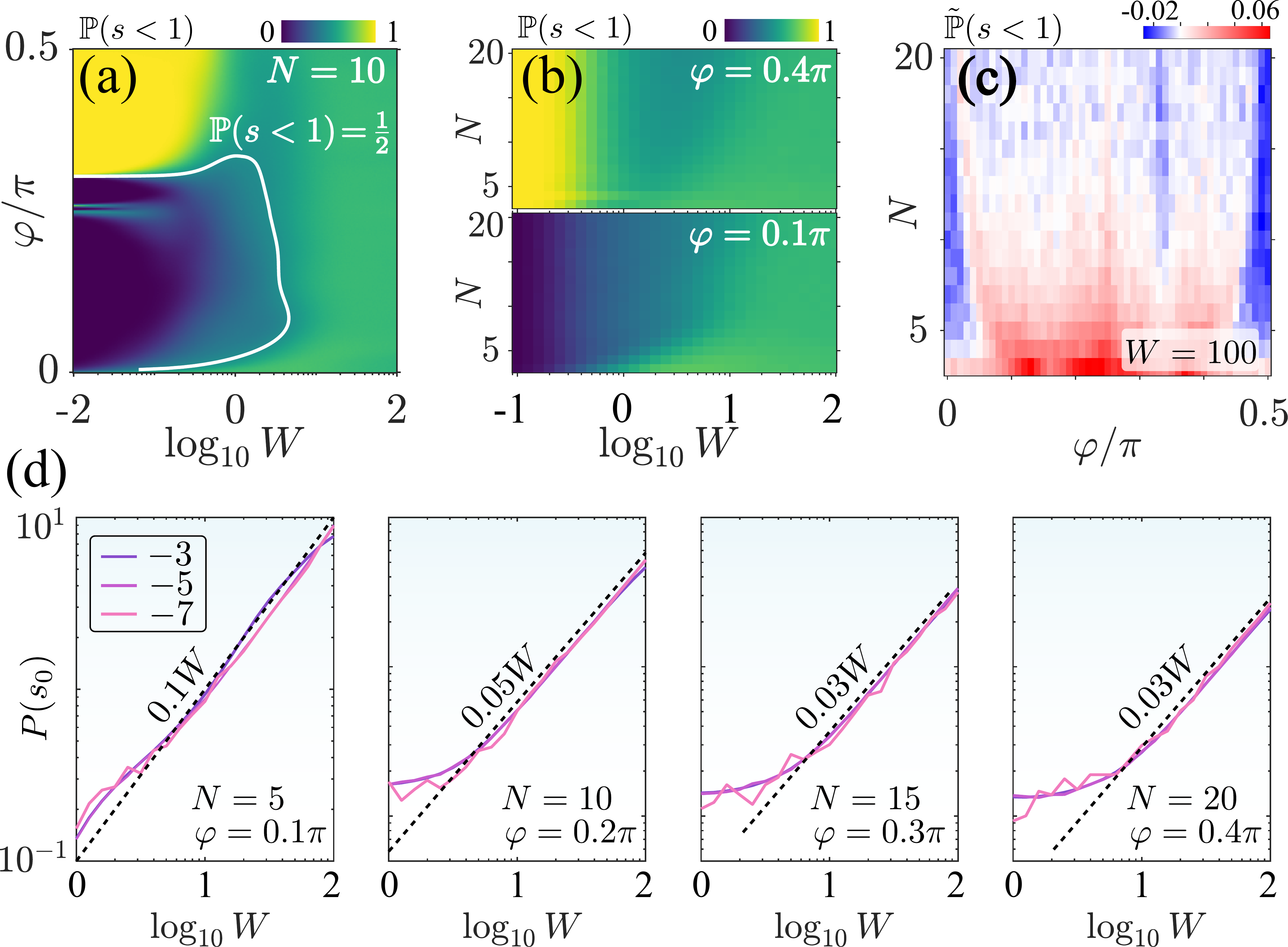}
		\caption{Correlation statistics of the reflection output. $\mathbb{P}(s<1)$ versus $\varphi$ and $W$ in (a), and versus $N$ and $W$ in (b). The white solid curve in (a) denotes the regime where $\mathbb{P}(s<1)=1/2$. $N=10$ in (a). $\varphi=0.4\pi$ ($\varphi=0.1\pi$) in the top (bottom) of (b). (c) $\tilde{\mathbb{P}}(s<1)$ for different chain sizes and phases, with $\tilde{\mathbb{P}}(s<1)=\mathbb{P}(s<1)-2/3$. $W=100$ in (c). (d) $P(10^{-\mu})$ versus disorder strength. Solid lines in different colors correspond to different values of $\mu$, with $\mu=-3,\ -5,\ -7$. Dashed lines are the numerical slopes. In (d), we only consider the contributions from non-interacting transition paths, whose validation is discussed in\,\cite{supp}. Results are obtained from $10^{10}$ disorder realizations.}\label{fig5}
	\end{figure}

	\emph{Reflection photon correlations with system disorder}.---One can expect that correlations in the reflection output exhibit behaviors distinct from those obtained in the transmission output, since reflected photons necessarily interact with the qubits, whereas transmitted photons can pass through the waveguide without interaction. In the presence of disorder, it is evident that $g_{\rm R}=0$ for $N=1$, implying that only PPB events occur. For a two-qubit chain, the correlation function is given by $g_{\rm R}=|(-i+i\exp(2i\varphi)+2\Delta_1+2\Delta_2)(\exp(2i\varphi)+(2\Delta_1-i)(2\Delta_2-i))/((\Delta_1+\Delta_2-i)(2\Delta_2-i+\exp(2i\varphi)(2\Delta_1+i))^2)|^2$. Based on this explicit expression, in the Supplemental Materials\,\cite{supp}, we show analytically and numerically that
	\begin{align}\label{eq6}
		\mathbb{P}(s<1)\ge\frac{2}{3},\ \ \ \ P(0)=0.
	\end{align}
	This result reveals two features: (i) Introducing disorder enables PA events, with a probability exceeding $1/2$, indicating that the output light is more likely to be antibunched; (ii) Despite the possibility of PA, the two-qubit system rules out the possibility of PPB, due to the absence of destructive interference of two-photon scattering paths. In fact, only strong PA can occur, provided that one qubit is far off-resonant while the other remains nearly resonant. In this case, the two-photon scattering probability amplitude $\langle {\rm \phi}^2_-|\psi^2\rangle$ is near zero, due to a large detuning, $|\Delta_1+\Delta_2|\gg1$, between the two-excitation state of the qubits and the two-photon state of the input; meanwhile, the single-photon scattering probability amplitude $\langle {\rm \phi}^1_-|\psi^1\rangle$ is near unity, since the nearly resonant qubit blocks the transmission of single photons from the input. However, the PPB events can never occur, since $\langle {\rm \phi}^2_-|\psi^2\rangle$ can never be exactly zero, due to the finite detuning of the far off-resonant qubit.
	
	For a chain with $N>2$, the behavior of $\mathbb{P}(s<1)$ is distinct from those obtained in chains with fewer qubits. In Figs.\,\ref{fig5}(a,b), we present the results of $\mathbb{P}(s < 1)$ for different system parameters. Our results reveal that, for $W\ll1$, the system exhibits a high (low) probability of PA when $0.3\pi<\varphi<0.5\pi$ ($0<\varphi<0.25\pi$). As disorder strength increases, the probability of PA decreases (increases) when $0.3\pi<\varphi<0.5\pi$ ($0<\varphi<0.25\pi$). In the strong-disorder regime, $\mathbb{P}(s < 1)$, especially for large chain sizes, saturates at a value close to $2/3$ [see Fig.\,\ref{fig5}(c)]. This value corresponds to the probability of PA for the two-qubit system with $\varphi=0$\,\cite{supp}. Regarding the probability of NPPB, its value increases and follows a power-law scaling with disorder strength when $W\gg1$ [see Fig.\,\ref{fig5}(d)]. This is in contrast to the results of the transmission output, where
	$P(10^{-3})$ rapidly decreases when disorder becomes too strong. We show in the Supplemental Materials that\,\cite{supp}, the enhancement of the probability of NPPB events in the reflection output stems from the fact that NPPB events involve solely the nearly completely destructive interference of two-photon non-interacting scattering paths.
	
	\emph{Conclusion}.---In summary, we have investigated the possibility of generating repulsively correlated photons in a chain of qubits coupled to a 1D waveguide. We found that, in the presence of disorder, it is possible to produce antibunched and nearly perfect blockaded photons, provided that the system parameters $N$, $\varphi$, and $W$ are suitably chosen. Furthermore, we demonstrate that the interplay among these parameters can significantly modulate the probability of strongly correlated photon events. Notably, our results reveal that the probabilities of PA and NPPB increase with the number of qubits according to a power-law scaling, while the probability of NPPB exhibits a power-law scaling with disorder strength. Our findings not only reveal the critical role of disorder in enabling strong photon correlations but also pave the way for disorder-engineered strongly correlated photons and potential single-photon sources.
	
	\emph{Acknowledgement}.---We sincerely thank Prof. Francesco Ciccarello for his valuable suggestions and insightful comments, which have greatly improved this work. We also thank Zhi-Guang Lu for his valuable suggestions. X.-Y. L. is supported by the National Science Fund for Distinguished Young Scholars of China (Grant No. 12425502), the Innovation Program for Quantum Science and Technology (Grant No. 2024ZD0301000), the National Key Research and Development Program of China (Grant No. 2021YFA1400700), and the Fundamental Research Funds for the Central Universities (Grant No. 2024BRA001). F.N. is supported in part by: the Japan Science and Technology Agency (JST) [via the CREST Quantum Frontiers program Grant No. JPMJCR24I2, the Quantum Leap Flagship Program (Q-LEAP), and the Moonshot R\&D Grant Number JPMJMS2061], and the Office of Naval Research (ONR) Global (via Grant No. N62909-23-1-2074).
	
	\emph{Data availability}.---The data that support the findings of this Letter are not publicly available. The data are available from the authors upon reasonable request.

	\let\oldaddcontentsline\addcontentsline
	\renewcommand{\addcontentsline}[3]{}
	%
	

\let\addcontentsline\oldaddcontentsline
\onecolumngrid

\renewcommand{\thefigure}{S\arabic{figure}}
\setcounter{figure}{0}
\setcounter{figure}{0}
\renewcommand{\theequation}{S\arabic{equation}}
\setcounter{equation}{0}
\renewcommand{\thetable}{S\arabic{table}}
\setcounter{table}{0}

\newpage

\setcounter{page}{1}\setcounter{secnumdepth}{3} \makeatletter
\begin{center}
	{\Large \textbf{ Supplemental Material for \\
			``Disorder-Induced Strongly Correlated Photons in Waveguide QED''}}
\end{center}


%

	\title{Supplemental Material for ``Disorder-induced Strongly Correlated Photons in Waveguide QED''}
	\author{Guoqing Tian}
	\affiliation{School of Physics and Institute for Quantum Science and Engineering, Huazhong University of Science and Technology, and Wuhan institute of quantum technology, Wuhan, 430074, P. R. China}
	\author{Li-Li Zheng}
	\affiliation{School of Artificial Intelligence, Jianghan University, Wuhan 430074, China}
	\author{Zhi-Ming Zhan}
	\affiliation{School of Artificial Intelligence, Jianghan University, Wuhan 430074, China}
	\author{Franco Nori}
	\affiliation{Quantum Computing Center, RIKEN, Wakoshi, Saitama 351-0198, Japan}
	\affiliation{Department of Physics, The University of Michigan, Ann Arbor, Michigan 48109-1040, USA}
	\author{Xin-You L\"{u}}\email{xinyoulu@hust.edu.cn}
	\affiliation{School of Physics and Institute for Quantum Science and Engineering, Huazhong University of Science and Technology, and Wuhan institute of quantum technology, Wuhan, 430074, P. R. China}
	
	\date{\today}
	\maketitle
	This supplemental materials contain eight parts: I. A detailed derivation of the second-order equal-time correlation function; II. Analytical calculations of the probability of photon antibunching (PA) and perfect photon blockade (PPB) for few-qubit systems; III. The physical mechanism of nearly perfect photon blockade (NPPB); IV. Correlation functions in weak- and strong-disorder limit; V. Effects of losses on non-waveguide modes; VI. Effects of chirality in coupling to waveguide modes; VII. Effects of finite bandwidth of input state; VIII. The details of numerical method used in this work.
	\tableofcontents
	
	\newpage

	\section{Derivation of second-order correlation function}
	In this section, we present the Gorini-Kossakowski-Sudarshan-Lindblad master equation for our setup and, by incorporating the input-output formalism, derive the formal expressions for the correlation functions in the transmission and reflection outputs. To this end, we consider a one-dimensional array of $N$ qubits coupled to a single optical waveguide bath. Under the rotating-wave approximation, the time evolution of the system is governed by the Hamiltonian
	\begin{align}\label{eq1}
		H=\sum_{m=1}^N\omega_m\sigma^{\dagger}_m\sigma_m+\sum_{\mu={\rm R},{\rm T}}\int_{-\infty}^{\infty}{\rm d}\omega \ \omega a^{\dagger}_{\mu}(\omega)a_{\mu}(\omega)+
		\sum_{m,\mu}\int_{-\infty}^{\infty}{\rm d}\omega\  (\kappa_{m,\mu}(\omega)a^{\dagger}_{\mu}(\omega)\sigma_m+h.c.).
	\end{align}
	Here, $\omega_m=\omega_0+\Delta_m$ denotes the transition frequency of the $m$th qubit, and $\sigma_m=|g_m\rangle\langle e_m|$ is its corresponding lowering operator. The operator $a_{\mu}(\omega)$ is the boson annihilation operator for a bath mode with frequency $\omega$, where the subscript $\mu={\rm T}$ (transmission) or R (reflection) distinguishes the mode. We assume weak inhomogeneity, namely, $|\Delta_m|\ll \omega_0$. In this regime, the coupling strengths satisfy $\kappa_{m,\rm T/R}(\omega)=\mathrm{g}_{m,{\rm,T/R}} \exp((-/+)i\omega x_m/c)\approx \mathrm{g_{T/R}} \exp((-/+)i\omega x_m/c)$, so that the interaction strength for each qubit is approximately homogeneous, i.e., $|\kappa_{m,\rm T/R}(\omega)|\approx \mathrm{g_{T/R}}$. The qubits are assumed to be equally spaced with separation $d$, so that $x_m=md$. For simplicity, we set $v_g=1$.
	
	We first derive the master equation describing the time evolution of the system, which further supports the calculation of the field correlation function. The derivations mainly follow those in\,\cite{caneva2015quantum_supp,asenjo-garcia_exponential_2017_supp,sheremet_waveguide_2023_supp}. Following standard derivations, the Heisenberg equation for the field operator are obtained as
	\begin{align}
		\dot{a}_{\mathrm{T}}(\omega,t)=-i\omega a_{\mathrm{T}}(\omega,t)-i\mathrm{g_{T}}e^{-i\omega x_m}\sigma_m(t),\ \ \ \ \dot{a}_{\mathrm{R}}(\omega,t)=-i\omega a_{\mathrm{R}}(\omega,t)-i\mathrm{g_{R}}e^{i\omega x_m}\sigma_m(t),
	\end{align}
	which can be formally integrated as
	\begin{equation}
		\begin{aligned}
			&a_{\rm T}(\omega,t)=e^{-i\omega (t-t_0)}a_{\rm T}(\omega,t_0)-i{\rm g_T}\sum_m\int_{t_0}^t e^{-i\omega(t-\tau)-i\omega x_m}\sigma_m(\tau)\dd{\tau},\ \ a_{\rm R}(\omega,t)=-i{\rm g_R}\sum_m\int_{t_0}^t e^{-i\omega(t-\tau)+i\omega x_m}\sigma_m(\tau)\dd{\tau},
		\end{aligned}
	\end{equation}
	where we have assumed that photons are injected from the left side of the waveguide, such that $a_{\rm R}(\omega,t_0)=0$. Consequently, the Heisenberg equations for the atomic operator are given by
	\begin{align}
		\dot{\sigma}_m(t)=-i\omega_m\sigma_m(t)-i\sigma^z_m(t)\qty(\int_{-\infty}^{\infty}{\rm{g_T}}a_{\rm T}(\omega,t)e^{i\omega x_m}\dd{\omega}+\int_{-\infty}^{\infty}{\rm{g_R}}a_{\rm R}(\omega,t)e^{-i\omega x_m}\dd{\omega}).
	\end{align}
	Substituting the solutions of the field operators into the equations for the atomic operator, we have
	\begin{equation*}
		\begin{aligned}
			\dot{\sigma}_m(t)&=-i\omega\sigma_m(t)-i\sigma^z_m(t)\Bigg({\rm g_T}\int_{-\infty}^{\infty}e^{-i\omega(t-t_0)+i\omega x_m}a_{\rm T}(\omega,t_0)\dd{\omega} \\ &-i{\rm g^2_T}2\pi \sum_n \Theta(x_m-x_n)\sigma_n(t-|x_m-x_n|)-i{\rm g^2_R}2\pi \sum_n \Theta(x_n-x_m)\sigma_n(t-|x_m-x_n|)\Bigg) \\
			&\approx-i\omega\sigma_m(t)-i\sigma^z_m(t)\Bigg({\rm g_T}\int_{-\infty}^{\infty}e^{-i\omega(t-t_0)+i\omega x_m}a_{\rm T}(\omega,t_0)\dd{\omega} \\ &-i{\rm g^2_T}2\pi \sum_n \Theta(x_m-x_n)\sigma_n(t)e^{i\omega_0|x_m-x_n|}-i{\rm g^2_R}2\pi \sum_n \Theta(x_n-x_m)e^{i\omega_0|x_m-x_n|}\sigma_n(t)\Bigg).
		\end{aligned}
	\end{equation*}
	In the last line, we have used the Markovian approximation
	\begin{align}
		\sigma_n(t-|x_m-x_n|/c)\approx \sigma_n(t)e^{i\omega_n|x_m-x_n|/c}=\sigma_n(t)e^{i\omega_0(1+\Delta_n/\omega_0)|m-n|d/c},
	\end{align}
	and, for our purposes, we neglect the dependence of $\Delta_n$ in the phase factor, i.e.,
	\begin{align}
		\sigma_n(t)e^{i\omega_n|x_m-x_n|/c}\approx\sigma_n(t)e^{i\omega_0|m-n|d/c}.
	\end{align}
	This approximation is valid when the condition $|\Delta_n|/\omega_0\ll N^{-1}$ holds. Equivalently, this condition is satisfied if $\mathcal{C}W\ll\omega_0/N$, where $\mathcal{C}\sim \order{1}$ is a dimensionless constant. Specifically, for a detuning $\Delta_n$ drawn from a normal distribution with standard deviation $W$, the inequality $|\Delta_n|\le \mathcal{C}W$ is met with probability $\erf(\mathcal{C}/\sqrt{2})$. For instance, taking $\mathcal{C}=2$ yields $\erf(\mathcal{C}/\sqrt{2})\approx 0.95$, implying that the condition $|\Delta_n|\le \mathcal{C}W$ is satisfies with near-unit probability. Thus, the approximation $|\Delta_n|/\omega_0\ll N^{-1}$ holds with probability near unit (0.95), provided that $\mathcal{C}W\ll\omega_0/N$ is met.
	
	Integrating the bath degrees of freedom, we have
	\begin{equation*}
		\begin{aligned}\label{equ_atom}
			\dot{\sigma}_m(t)&=-i\omega_m\sigma_m(t)-i\sqrt{\gamma_{\rm T}}\sigma^z_m(t)f(t-t_0,x_m) \\ &-\gamma_{\rm T}\sigma^z_m(t)\sum_n\sigma_n(t)\Theta(x_m-x_n)e^{i\omega_0|x_m-x_n|}-\gamma_{\rm R}\sigma^z_m(t)\sum_n\sigma_n(t)\Theta(x_n-x_m)e^{i\omega_0|x_m-x_n|},
		\end{aligned}
	\end{equation*}
	where $\gamma_{\rm T/R}=2\pi {\rm g^2_{T/R}}$ and $f(\tau,z)$ is defined by
	\begin{align}
		f(\tau,z)=\frac{1}{\sqrt{2\pi}}\int_{-\infty}^{\infty}e^{-i\omega(\tau-z)}\Tr\qty[\rho_E(t_0)a_{\rm T}(\omega,t_0)]\dd{\omega}.
	\end{align}
	Eq.\,(\ref{equ_atom}) coincides with the equation of motion governed by the master equation
	\begin{align}\label{ME}
		\dot{\rho}=-i((H_{\rm eff}+H_d)\rho-\rho(H^{\dagger}_{\rm eff}+H_d))+\sum_{mn}2\qty(\gamma_{\rm T}\Theta(m-n)\cos(|m-n|\varphi)+\gamma_{\rm R}\Theta(n-m)\cos(|m-n|\varphi))\sigma_m\rho\sigma^{\dagger}_n,
	\end{align}
	with
	\begin{align}\label{eq5}
		H_{\rm eff}=\sum_{m=1}^{N}\qty(\Delta_m-\frac{i(\gamma_{\rm nw}+\gamma_{\rm T}+\gamma_{\rm R})}{2})\sigma^{\dagger}_m\sigma_m    -i\sum_{m>n}^{N}\qty(\gamma_{\rm T}e^{i|m-n|\varphi}\sigma^{\dagger}_m\sigma_n+\gamma_{\rm R}e^{i|m-          n|\varphi}\sigma^{\dagger}_n\sigma_m),
	\end{align}
	and
	\begin{align}
		H_d=\sum_{m}\sqrt{\gamma_{\rm T}}(f(t-t_0,md)\sigma^{\dagger}_m+h.c.),
	\end{align}
	where $\varphi=\omega_0d/v_g$.
	
	The zero-time second-order photon correlations of the emitted field are defined as
	\begin{align}\label{g2_def}
		g_{\mu}=\frac{\langle a_{\mu,\rm{out}}^{\dagger}(t)a_{\mu,\rm{out}}^{\dagger}(t)a_{\mu,\rm{out}}(t)a_{\mu,\rm{out}}(t)\rangle}{\langle a_{\mu,\rm{out}}^{\dagger}(t)a_{\mu,\rm{out}}(t)\rangle^2},
	\end{align}
	where $\langle\bullet\rangle$ denotes the expectation value over the emitted field. The input-output relations are given by
	\begin{align}\label{in_out}
		a_{\rm T,\rm{out}}(t)=a_{\rm T,\rm in}(t)-i\sqrt{\gamma_{\rm T}}\sum_m e^{-im\varphi}\sigma_m(t),\ \ \ \ \ a_{\rm R,\rm{out}}(t)=-i\sqrt{\gamma_{\rm R}}\sum_m e^{im\varphi}\sigma_m(t),
	\end{align}
	with
	\begin{align}
		a_{\rm T,\rm in}(t)=\frac{1}{\sqrt{2\pi}}\int_{-\infty}^{\infty}e^{-i\omega(t-t_0)}a_{\rm T}(\omega,t_0)\dd{\omega}.
	\end{align}
	Let us now consider a right-propagating coherent pulse as an input (drive), given by $\rho_E(t_0)\propto {\rm exp}(\alpha a^{\dagger}_{\rm T}(\omega_{\rm in})-\alpha a_{\rm T}({\omega_{\rm in}}))|0\rangle$, where the input pulse is assumed to be resonant with the transition frequency for the ordered system, i.e., $\omega_{\rm in}=\omega_0$. In the long-time limit, the properties of the emitted field are fully determined by the steady state $\rho_{\rm ss}$ of the qubits (in the rotating frame with respect to $H_0=\omega_0\sum_m \sigma^{\dagger}\sigma_m$). By substituting the input-output relations into the definition of $g_{\mu}$, one obtains
	\begin{align}
		g_{\rm \mu}=\frac{\Tr(\rho_{\rm ss}C_{\mu}^{\dagger}C_{\mu}^{\dagger}C_{\mu}C_{\mu})}{\Tr(\rho_{\rm ss}C_{\mu}^{\dagger}C_{\mu})^2},
	\end{align}
	where $C_{\rm R}=-i\sqrt{\gamma_{\rm R}}\sum_{m}e^{i\varphi m}\sigma_m$ and $C_{\rm T}=\alpha-i\sqrt{\gamma_{\rm T}}\sum_{m}e^{-i\varphi m}\sigma_m$, $\rho_{\rm ss}$ denotes the steady state corresponding to the master equation Eq.\,(\ref{ME}) with $f(t-t_0,x_m)=\alpha e^{-i\omega_0 (t-t_0)}e^{im\varphi}$. In the weak-input limit ($\alpha\ll1$), one can solve for $\rho_{\rm ss}$ by neglecting quantum jumps, which yields the expansion $\rho_{\rm ss}\approx |G\rangle+\alpha|\psi^1\rangle+\alpha^2|\psi^2\rangle+\order{\alpha^3}$\,\cite{giant}. Here, $|G\rangle=\bigotimes_{m=1}^N|g_m\rangle$, $|\psi^1\rangle=-(H_{\rm eff}^{(1)})^{-1}H_+|G\rangle$ and $\ |\psi^2\rangle=-(H_{\rm eff}^{(2)})^{-1}H_+|\psi^1\rangle$ are the single- and two-excitation components of the truncated steady-state, respectively, with $H_+=\sqrt{\gamma_{\rm T}}\sum_{m} e^{im\varphi}\sigma^{\dagger}_m$. The operators $H_{\rm eff}^{(1)}$ and $H_{\rm eff}^{(2)}$ denote the single- and two-excitation subspace of the effective Hamiltonian $H_{\rm eff}$, respectively. After some calculations, one obtains for the transmitted field
	\begin{align}\label{eq9}
		g_{\rm T}=\frac{|1-2i\langle G|\tilde{C}_{\rm T}|\psi^1\rangle-\langle G|\tilde{C}_{\rm T}\tilde{C}_{\rm T}|\psi^2\rangle|^2}{|1-i\langle G|\tilde{C}_{\rm T}|\psi^1\rangle|^4}+\order{|\alpha|^2},
	\end{align}
	and for the reflected field
	\begin{align}\label{eq10}
		g_{\rm R}=\frac{\langle \psi^2|(\tilde{C}^{\dagger}_{\rm R})^2(\tilde{C}_{\rm R})^2|\psi^2\rangle}{\langle \psi^1|\tilde{C}^{\dagger}_{\rm R}\tilde{C}_{\rm R}|\psi^1\rangle^2}+\order{|\alpha|^2},
	\end{align}
	where $\tilde{C}_{\rm T}=\sqrt{\gamma_{\rm T}}\sum_{m}e^{-i\varphi m}\sigma_m$ and $\tilde{C}_{\rm R}=\sqrt{\gamma_{\rm R}}\sum_{m}e^{i\varphi m}\sigma_m$. These expressions can be further simplified as
	\begin{align}\label{gt_intef}
		g_{\rm T}=\frac{|1-2i\langle {\phi^1_{+}}|\psi^1\rangle-\langle {\phi^2_{\rm +}}|\psi^2\rangle|^2}{|1-i\langle {\phi^1_{+}}|\psi^1\rangle|^4},
	\end{align}
	and
	\begin{align}\label{gr_intef}
		g_{\rm R}=\frac{|\langle {\phi^2_{\rm -}}|\psi^2\rangle|^2}{|\langle {\phi^1_{\rm -}}|\psi^1\rangle|^4},
	\end{align}
	where $|{\phi^1_{\pm}}\rangle=\sqrt{\gamma_{\rm T}}\sum_m \exp(\pm im\varphi)\sigma^{\dagger}_m|G\rangle$ and $|{\phi^2_{\pm}}\rangle=2\gamma_{\rm T}\sum_{m>n} \exp(\pm i(m+n)\varphi)\sigma^{\dagger}_m\sigma^{\dagger}_n|G\rangle$. When the coupling strengths to left- and right-going modes are homogeneous, i.e., $\gamma_{\rm T}=\gamma_{\rm R}=\gamma/2$, these equations, Eqs.\,(\ref{gt_intef},\ref{gr_intef}), recover Eq.\,(2) of the main text.
	
\begin{figure*}
		\centering
		\includegraphics[width=14cm]{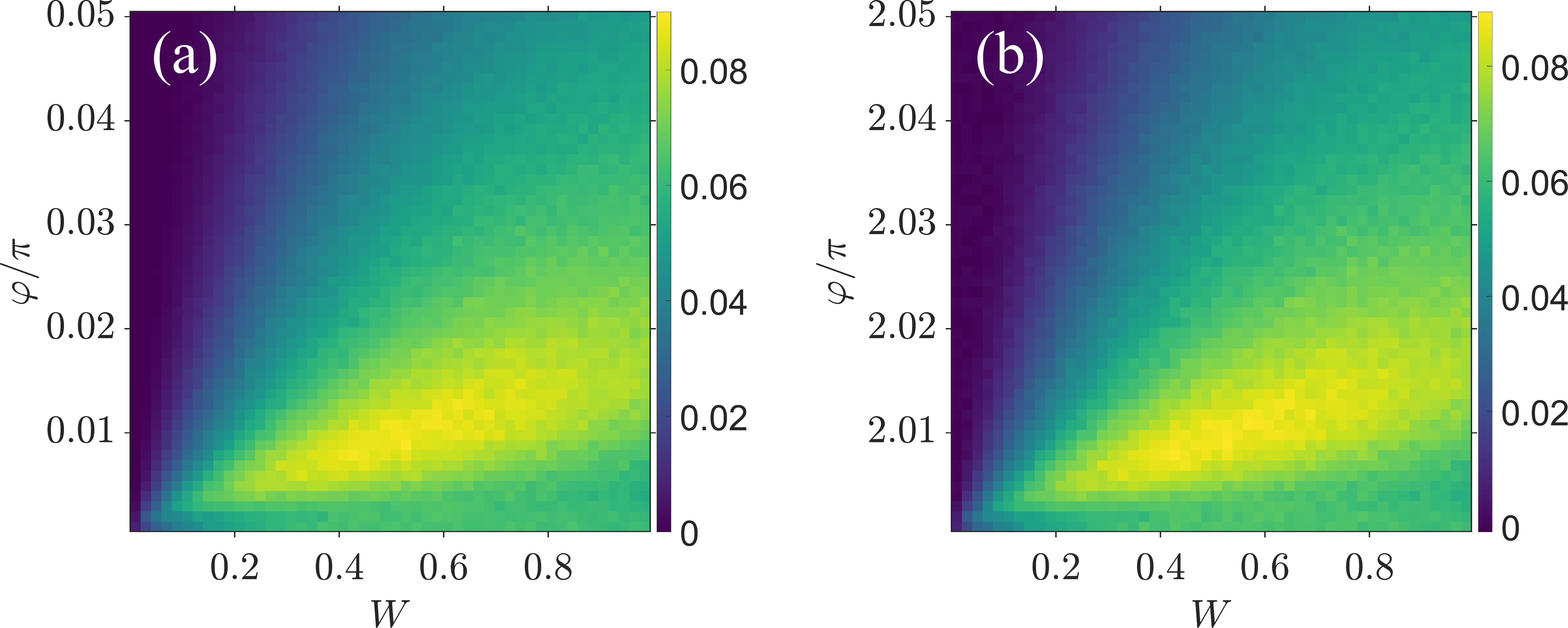}
		\caption{(a) Probability of PA versus $\varphi$ and $W$, where $\varphi$ ranges from $0.001\pi$ to $0.05\pi$. (b) Probability of PA versus $\varphi$ and $W$, where $\varphi$ ranges from $2.001\pi$ to $2.05\pi$}.\label{fig0}
	\end{figure*}

	We also emphasize the periodic nature of all the above derivations, including the master equation and the input-output relations, in the phase $\varphi$. Consequently, photon correlation should also be periodic in $\varphi$, with a period of $2\pi$. Likewise, all the statistic quantities studied in the main text, such as the probability of PA and the probability density function, are also periodic in $\varphi$ with period of $2\pi$. For example, the probability of PA in the transmission for $N=10$ (the inset of Fig.\,2(a) in the main text) with $\varphi/\pi\in\qty(0.001,0.05)$, and that with $\varphi/\pi\in\qty(2.001,2.05)$, are shown in Fig.\,\ref{fig0}. The results show good agreements with each other.
	
	
	\section{Analytical Calculations of $\mathbb{P}(s<1)$ and $P(0)$ for few-qubit systems.}
	In this section, we present detailed calculations for the probabilities of PA and PPB for few-qubit systems, which further substantiate the results reported in the main text. The calculations for the transmission output and for the reflection output with $\varphi=0$ are carried out analytically, whereas those for the reflection output with $\varphi\ne0$ are performed numerically.
	\subsection{Transmission}
	We first consider the single-qubit case. In this system, all two-photon scattering processes vanish, meaning that any terms involving $|\psi^2\rangle$ (or $\langle\psi^2|$) drop out. Since $H^{(1)}_{\rm eff}$ is a $1\times1$ matrix in this case, the evaluation of $|\psi^1\rangle$ is straightforward. After some algebraic manipulations, the second-order correlation function is given by
	\begin{align}\label{g2_1atom}
		g_{\rm T}=\frac{(1+4\Delta^2_1)^2}{16\Delta^2_1}.
	\end{align}
	Moreover, an analytical expression for the probability density function (PDF) corresponding to Eq.\,(\ref{g2_1atom}) can be obtained. Specifically, the PDF is calculated as
	\begin{align}\label{eq12}
		P(s)=\frac{1}{\sqrt{2\pi}W}\int_{-\infty}^{\infty} \delta\qty(\frac{(1+4\Delta^2_{1})^2}{16\Delta^4_{1}}-s)e^{-\Delta^2_1/2W^2}\ {\rm d}\Delta_1.
	\end{align}
	Here, the Dirac delta function is handled using the standard identity $\delta(f(x))=\sum_{j}\delta(x-x_j)/|f'(x_j)|$. After some calculations, one finds that $P(s)=0$ for $s<1$. Thus, we have
	\begin{align}
		\mathbb{P}(s<1)=P(0)=0.
	\end{align}
	For $s\ge1$, the PDF is given by
	\begin{align}
		P(s)=\frac{1}{4\sqrt{2\pi}W}\frac{1}{(\sqrt{s}-1)^{3/2}\sqrt{s}}e^{1/(1-1\sqrt{s})8W^2}.
	\end{align}
	The asymptotic behaviors for $P(s>1)$ are
	\begin{align}
		P(s\to 1^+)\sim \frac{1}{4\pi W}(s-1)^{-3/2}e^{-1/(4W^2(s-1))},\ \ \ \ \ P(s\to +\infty)\sim \frac{1}{4\sqrt{2}\pi W}s^{-5/4}.
	\end{align}
	Furthermore, the mode of the PDF, corresponding to the most probable value of the correlation function, is located at
	\begin{align}
		s_{\rm max}=\frac{1+56W^2+464W^4+\sqrt{W^8(1+28W^2)^2(1+56W^2+464W^4)}}{800W^4},
	\end{align}
	This result shows that the mode scales as $s_{\rm max}\sim W^{-4}$ for $W\to 0^+$ and $s_{\rm max}\sim 1$ for $W\to+\infty$. In other words, when $W\ll1$ the most probable output is strongly bunched, while for $W\gg1$ it is nearly coherent.
	
	For a two-qubit system, although the analytical expression become more involved, the calculation remains tractable. In this case, the correlation function is given by
	\begin{align}
		g_{\rm T}=\frac{(8\Delta^2_1\Delta^2_2+f_+-\cos(2\varphi))\qty(8\Delta^2_1\Delta^2_2(1+(\Delta_1+\Delta_2)^2)+(\Delta_1+\Delta_2)^2(f_{-}-\cos(\varphi))+4\Delta_1\Delta_2(\Delta_1+\Delta_2)\sin(2\varphi))}{64\Delta^4_1\Delta^4_2(1+(\Delta_1+\Delta_2)^2)},
	\end{align}
	with $f_{\pm}=1+2\Delta^2_1+2\Delta^2_2\pm4\Delta_1\Delta_2\cos(2\varphi)-2(\Delta_1+\Delta_2)\sin(2\varphi)$. It is practically infeasible to derive a closed-form expression for the corresponding PDF in this case. Nonetheless, since the correlation function $g_{\rm T}\ge1$, we have $\mathbb{P}(s < 1) = 0$ and $P(0) = 0$.
	
	\subsection{Reflection}
	For the single-qubit system, the absence of two-photon scattering processes implies that $\langle {\phi^2_{\rm -}}|\psi^2\rangle=0$, so that $g_{\rm R}=0$. For the two-qubit system, we first consider the case $\varphi=0$. In this situation, the correlation function reduces to
	\begin{align}
		g_{\rm R} = \frac{(\Delta_1+\Delta_2)^2+4\Delta^2_1\Delta^2_2}{(\Delta_1+\Delta_2)^2+(\Delta_1+\Delta_2)^4}.
	\end{align}
	The probability of PA is then calculated from
	\begin{align}\label{eq19}
		\mathbb{P}(s<1)=\int_0^1 P(s){\rm d}s =\frac{1}{2\pi W^2}\int_{-\infty}^{\infty}{\rm d}\Delta_1\int_{-\infty}^{\infty}{\rm d}\Delta_2\ \exp(-\frac{\Delta^2_1+\Delta^2_2}{2W^2})\Theta(g_{\rm R}-1),
	\end{align}
	where $\Theta(x)$ denotes the Heaviside step function. Changing the variables as $2\Omega=\Delta_1+\Delta_2$ and $2\Lambda=\Delta_1-\Delta_2$, one obtains $g_{\rm R}=(\Omega^2+(\Omega^2-\Lambda^2)^2)/(\Omega^2+4\Omega^4)$. PA occurs when
	\begin{align}
		\abs{\Omega}>\frac{1}{\sqrt{3}}\abs{\Lambda}.
	\end{align}
	It follows that the probability of PA is given by
	\begin{align}
		\mathbb{P}(s<1)=\frac{1}{\pi W^2}\iint_{\abs{\Omega}>\frac{\abs{\Lambda}}{\sqrt{3}}}{\rm d}\Omega{\rm d}\Lambda\ \exp(-\frac{\Omega^2+\Lambda^2}{W^2})
		=\frac{2}{3},
	\end{align}
	which recovers the equality stated in Eq.(4) of the main text. To demonstrate the result $P(0)=0$ presented in the main text, we now analyze the asymptotic behavior of $P(s)$ for $s\ll1$. Starting from
	\begin{equation}\label{eq22}
		P(s)=\frac{4}{\pi W^2}\int_{0}^{\infty}{\rm d}\Omega\int_{0}^{\infty}{\rm d}\Lambda\ \exp(-\frac{\Omega^2+\Lambda^2}{W^2})
		\delta\qty(\frac{\Omega^2+(\Omega^2-\Lambda^2)^2}{\Omega^2+4\Omega^4}-s).
	\end{equation}
	a change of variables $\Omega^2\to\Omega$ and $\Lambda^2\to\Lambda$ yields
	\begin{equation}
		P(s)=\frac{1}{\pi W^2}\int_{0}^{\infty}{\rm d}\Omega\int_{0}^{\infty}{\rm d}\Lambda\ \frac{1}{\sqrt{\Omega\Lambda}}\exp(-\frac{\Omega+\Lambda}{W^2})
		\delta\qty(\frac{\Omega+(\Omega-\Lambda)^2}{\Omega+4\Omega^2}-s).
	\end{equation}
	Carrying out the integral over $\Lambda$ leads to
	\begin{align}\label{B5}
		P(s)=\frac{1}{\pi W^2}\int_{(1-s)/4s}^{\infty}{\rm d}\Omega\ \exp(-\Omega/W^2)\frac{(1+4\Omega)}{\sqrt{4s\Omega+s-1}}
		\qty[\frac{\exp(\frac{-\Omega_+}{W^2})}{\sqrt{\Omega_+}}+\frac{\exp(\frac{-\Omega_-}{W^2})}{\sqrt{\Omega_-}}],
	\end{align}
	where $\Omega_{\pm}=\Omega\pm\sqrt{\Omega(4s\Omega+s-1)}$. In the limit $s\ll1$, the lower limit of the $\Omega$ integral can be approximated as $(1-s)/4s\approx1/4s$. In this regime, one can approximate $\Omega_{\pm}\approx\Omega$ and $\sqrt{4s\Omega+s-1}\approx\sqrt{4s\Omega-1}$. With these approximations, the expression simplifies to
	\begin{align}\label{ps_asymp}
		P(s)\approx\frac{2}{\pi W^2\sqrt{s}}\int_{1/4s}^{\infty}{\rm d}\Omega\ e^{-2\Omega/W^2}\frac{1}{\sqrt{4s\Omega-1}}.
	\end{align}
	This asymptotic form reveals that strong photon antibunching ($g_{\rm R}\ll1$) is associated with large detuning, as dictated by the lower integration bound $\Omega=\Delta_1+\Delta_2>1/4s\sim s^{-1}$. Performing the integral over $\Omega$ yields
	\begin{align}\label{eq15}
		P(s\ll1)\sim \frac{1}{\sqrt{2\pi}W}s^{-1}\exp(-\frac{1}{2W^2s}).
	\end{align}
	From this expression it is evident that as $s\to0$, $P(s)\to 0$; hence, $P(0)=0$, in agreement with the result stated in the main text. Finally, we compare this analytical approximation to numerical integration of the starting expression (Eq.\,(\ref{eq22})) [see Figs.\,\ref{fig1}(a-c)].
	
	\begin{figure*}
		\centering
		\includegraphics[width=18cm]{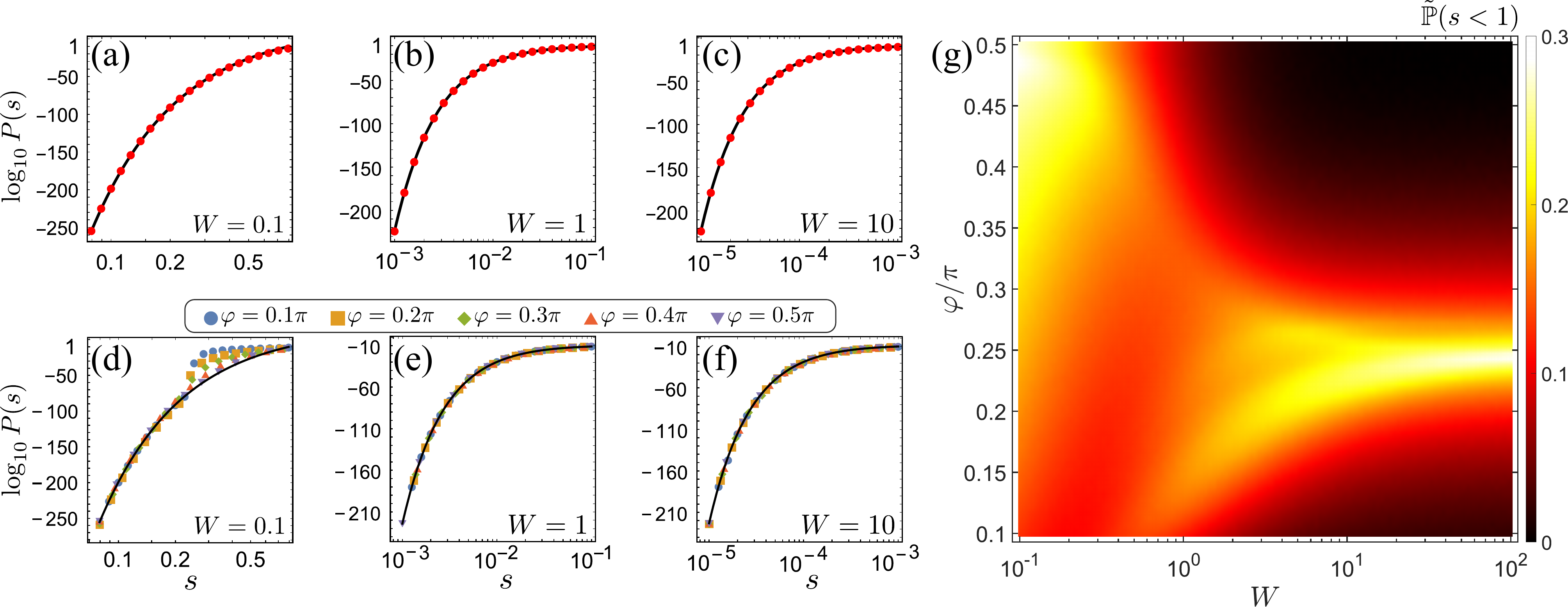}
		\caption{(a-c) The asymptotic behaviors of $P(s)$. The chosen disorder strengths are $W=0.1$ (a), $W=1$ (b), and $W=10$ (c). In all plots, red dots are obtained from the numerical integration of Eq.\,(\ref{eq22}), and black lines represent the analytical solutions from Eq.\,(\ref{eq15}). (d-f) The asymptotic behaviors of $P(s)$. The chosen disorder strengths are $W=0.1$ (d), $W=1$ (e), and $W=10$ (f). The different symbols correspond to systems with different values of $\varphi$. In all plots, the symbols are obtained from the numerical integration with the correlation expression replaced by Eq.\,(\ref{eq26}), and black lines represent the analytical solutions from Eq.\,(\ref{eq15}). (g) $\tilde{\mathbb{P}}(s<1)=\mathbb{P}(s<1)-2/3$ versus phase and disorder strength. The result is obtained from numerical integration of Eq.\,(\ref{eq19}),  while the expression of correlation is replaced by Eq.\,(\ref{eq26})}.\label{fig1}
	\end{figure*}

	For the system with $\varphi\ne0$, the correlation function is given by
	\begin{align}\label{eq26}
		g_{\rm R}=\abs{\frac{(-i+ie^{2i\varphi}+2\Delta_1+2\Delta_2)(e^{2i\varphi}+(2\Delta_1-i)(2\Delta_2-i))}{(\Delta_1+\Delta_2-i)(2\Delta_2-i+e^{2i\varphi}(2\Delta_1+i))^2}
		}^2.
	\end{align}
	In this case, the analytical forms for both $\mathbb{P}(s<1)$ and $P(0)$ are practically infeasible to obtain. Fig.\,\ref{fig1}(g) displays the numerical results for $\mathbb{P}(s<1)$, which now depend on both the phase and the disorder strength. Notably, the numerical data imply that $\mathbb{P}(s<1)$ is bounded from below by $2/3$, the value obtained in the Dicke limit for a two-qubit system. As the disorder strength increases, $\mathbb{P}(s<1)$ attains a maximum at $\varphi\sim 0.25\pi$, and then saturates to the lower bound of $2/3$ when $\varphi\sim 0$ or $0.5\pi$.
	
	Regarding $P(0)$, Figs.\,\ref{fig1}(d-f) illustrate the asymptotic behavior of $P(s)$ for $s\ll1$. It is evident that $P(s\ll1)$ displays essentially the same asymptotic form as in the case of $\varphi=0$. This similarity arises because strong PA is primarily associated with samples $\qty{\Delta_1,\Delta_2}$ in which one detuning $|\Delta_i|\ll1$ (i.e., nearly resonant), while the other $|\Delta_{j\ne i}|\gg1$ (i.e., far off-resonant). Physically, this scenario corresponds to one qubit interacting strongly with the input while the other is nearly transparent to it, effectively reducing the system to a single-photon absorber/emitter. However, since the detunings are sampled from a Gaussian distribution, the probability of obtaining $|\Delta_{j\ne i}|\gg1$ decreases exponentially with $|\Delta_{j\ne i}|$, which ultimately leads to an exponential decay of $P(s)$ as $s\to0$. Consequently, $P(0)=0$ is recovered in this regime.
	
	These findings demonstrate that, even when $\varphi\ne0$, the essential asymptotic behavior of the correlation function remains consistent with the $\varphi=0$ case, while the overall probability $\mathbb{P}(s<1)$ exhibits a nontrivial dependence on both the phase and the disorder strength.

	\section{Physical mechanism of NPPB}
	\begin{table}\label{t1}
		\centering
		\renewcommand{\arraystretch}{1.5}
		\setlength{\tabcolsep}{10pt}
		\begin{tabular}{|c|c|c|c|}
			\hline
			$g_{\rm T}$ & $\Delta_1$ & $\Delta_2$ & $\Delta_3$ \\
			\hline
			$10^{-8}$ & $0.149124450372206$ & $-0.053903424589490$ & $0.144085703957167$ \\
			\hline
			$10^{-10}$ & $0.148134188883455$ & $-0.055253952848190$ & $0.144762975264912$ \\
			\hline
			$10^{-12}$ & $0.149005562567387$ & $-0.054129160721382$ & $0.144182673551411$ \\
			\hline
		\end{tabular}
		\caption{Partial solutions $\qty{\Delta_1,\Delta_2,\Delta_3}$ of equation $g_{\rm T}=s_0$ with $s_0=10^{-8},10^{-10},10^{-12}$. Here $\varphi=0.04\pi$, which is the same with the parameters in Fig.\,1(c) of main text.}
	\end{table}
	\begin{figure*}
		\centering
		\includegraphics[width=18cm]{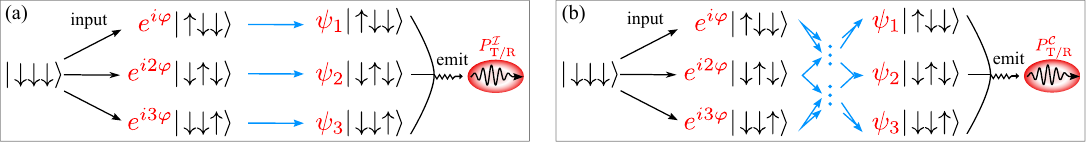}
		\caption{(a-b) Schematics of single-photon scattering path involves (a) non-interacting path (b) interacting path for an array with $N=3$.}\label{fig2}
	\end{figure*}
	\begin{figure*}
		\centering
		\includegraphics[width=18cm]{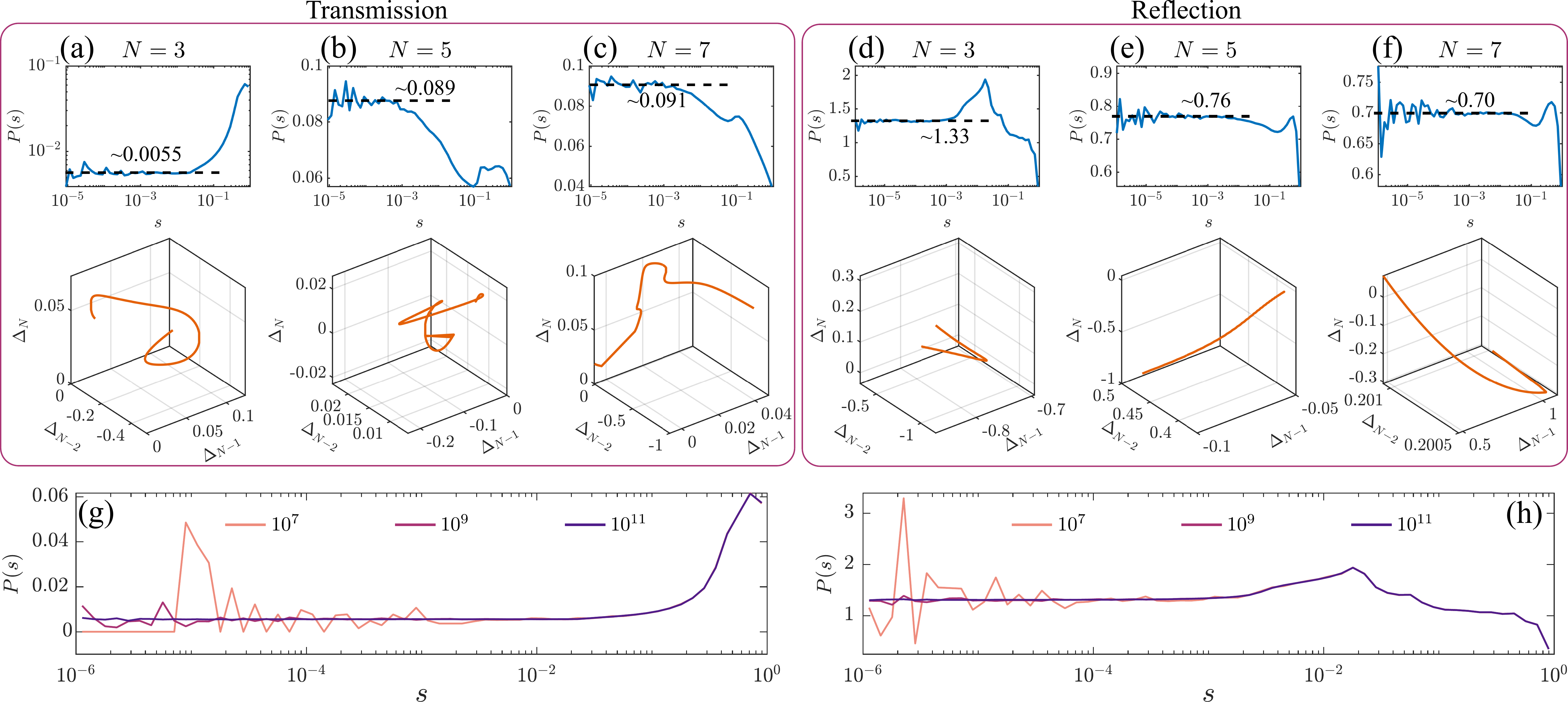}
		\caption{(a-c) Top: PDFs; Bottom: Solutions of $g_{\rm T}=10^{-10}$ for the transmission output. The chosen parameters are $\varphi=0.01\pi$ and $W=0.15$; the chain sizes are indicated above the plots. (d-f) Top: PDFs; Bottom: Solutions of $g_{\rm R}=10^{-10}$ for the reflection output. The chosen system parameters are $\varphi=0.5\pi$ and $W=1$; the chain sizes are indicated above the plots. In the top of (a-f), the results are obtained from $10^9$ disorder realizations. (g-h) PDFs for the transmission (g) and reflection (h) outputs under different numbers of disorder realizations. The chosen parameters are $N=3$, $\varphi=0.01\pi$, $W=0.15$ in (g) and $N=3$, $\varphi=0.5\pi$, $W=1$ in (h).}\label{fig3}
	\end{figure*}
	
	Although the solutions presented in the main text correspond to $g_{\rm T}=10^{-10}$, one can also obtain solutions for $g_{\rm T}=\epsilon$, where $\epsilon$ can be made arbitrarily close to 0 as computational precision increases. To achieve an NPPB event, i.e., $g_{\rm T/R}=\epsilon\to0$, the detunings of the qubits must be very finely tuned (see Table.\,\hyperref[t1]{S1} for an example). This requirement for fine-tuning of parameters for an NPPB event suggests that the underlying physical mechanism is the destructive interference of quantum paths\,\cite{sing_photon1_supp,sing_photon2_supp}.
	
	To manifest the interference effect, note that the photon correlations depend solely on the single- and two-photon scattering processes. Consequently, the overall quantum path comprises three contributions: (i) a single-photon scattering path with probability amplitude $\langle{\phi^1_{\pm}}|\psi^1\rangle$; (ii) a two-photon scattering path with probability amplitude $\langle{\phi^2_{\pm}}|\psi^2\rangle$; (iii) a free propagation path with probability amplitude $1$. From Eqs.\,(\ref{gt_intef}-\ref{gr_intef}), NPPB in the transmission
	output involves nearly completely destructive interference among the single-photon scattering path, the two-photon scattering path, and the free propagation path, whereas NPPB in the reflection output involves nearly completely destructive interference solely of the two-photon scattering path. Furthermore, the single- and two-photon scattering paths are determined by transition paths, which are governed by the non-Hermitian effective Hamiltonian $H_{\rm eff}$. These transition paths dictate how the qubits are excited by the input and how the emitted photons interfere with each other.
	
	More specifically, in the single-photon scattering path the fully inverted chain of qubits is first excited by the input to the state, $|G\rangle+\alpha H_+|G\rangle$. Then, the single-excitation component, $\alpha H_+|G\rangle\propto \sum_{m}\exp(im\varphi)\sigma^{\dagger}_m|G\rangle$, transitions to the steady state through an infinite number of transition paths. To see this, we can rewrite
	\begin{align}
		|\psi^1\rangle = -(H^{(1)}_{\rm eff})^{-1}H_+|G\rangle=\lim_{z\to 0}\frac{1}{z-H^{(1)}_{\rm eff}}H_+|G\rangle=\lim_{z\to 0}(G^{(1)}_{\mathcal{I}}(z)+G^{(1)}_{\mathcal{C}}(z))H_+|G\rangle,
	\end{align}
	with
	\begin{align}
		G^{(1)}_{\mathcal{I}}(z)=\frac{1}{z-H^{(1)}_0},\ \ \ \ \ G^{(1)}_{\mathcal{C}}(z)=\sum_{j=1}^{\infty}G^{(1)}_{\mathcal{C},j}(z),\ \ \ \ \ G^{(1)}_{\mathcal{C},j}(z)=\qty(G^{(1)}_{\mathcal{I}}(z)T^{(1)})^jG^{(1)}_{\mathcal{I}}(z).
	\end{align}
	Here $H_0=\sum_m(\Delta_m-i\gamma/2)\sigma^{\dagger}_m\sigma_m$ and $T=-i\gamma\sum_{m\ne n}\exp(i\varphi|m-n|)\sigma^{\dagger}_m\sigma_n/2$ represent the free and interaction terms of the non-Hermitian Hamiltonian, respectively; the superscript $(1)$ denotes the single-excitation sector. Now, assuming that the single-excitation component of the steady state can be written as $|\psi^1\rangle=\sum_{m}\psi_m\sigma^{\dagger}_m|G\rangle$, the unnormalized probability amplitude is given by $\psi_m=\psi^{\mathcal{I}}_{m}+\psi^{\mathcal{C}}_{m}$, with
	\begin{align}
		\psi^{\mathcal{I}}_{m}=\lim_{z\to 0}\langle G|\sigma_mG^{(1)}_{\mathcal{I}}(z)H_+|G\rangle=-\sqrt{\frac{\gamma}{2}}\frac{e^{im\varphi}}{\Delta_m-i\gamma/2},\ \ \ \ \psi^{\mathcal{C}}_{m}=-\sqrt{\frac{\gamma}{2}}\lim_{z\to 0}\sum_{n=1}^{N}\sum_{j=1}^{\infty}e^{in\varphi}\langle G|\sigma_mG^{(1)}_{\mathcal{C},j}(z)\sigma^{\dagger}_n|G\rangle.
	\end{align}
	The first term, $\psi^{\mathcal{I}}_m$, represents the steady-state probability amplitude of qubits individually interacting with the waveguide, while the second term, $\psi^{\mathcal{C}}_m$, represents the probability amplitude for qubits collectively interacting with the waveguide. This collective term involves transitions between different qubits mediated by the long-range interaction of photons. Such transitions can be viewed as an emission-reabsorption process, whereby photons emitted by one qubit are reabsorbed by another. Specifically, for the $n$-th qubit initially excited by the input, the excitation transfers to the $m$-th qubit through an infinite sequence of transitions, with the unnormalized probability amplitudes for the $j$-th path given by $e^{in\varphi}\sqrt{\gamma/2}\langle G|\sigma_mG^{(1)}_{\mathcal{C},j}(z)\sigma^{\dagger}_n|G\rangle$.
	
	After excitation, each excited qubit in the state $\psi_m\sigma^{\dagger}_m|G\rangle$ can emit a single photon into the waveguide with an (unnormalized) probability amplitude $\psi_m$. The propagation of this emitted photon acquires a phase factor $\exp(\pm im\varphi)$, where the minus (plus) sign corresponds to the transmission (reflection) output. Consequently, the final probability amplitude for the single-photon path is given by
	\begin{align}
		\langle{\rm sup^1_{\pm}}|\psi^1\rangle=P^{\mathcal{I}}_{\rm T/R}+P^{\mathcal{C}}_{\rm T/R},
	\end{align}
	with
	\begin{align}
		P^{\mathcal{I}}_{\rm T/R}=-\frac{\gamma}{2}\sum_{m=1}^{N}\frac{{\rm exp}[(0/2)im\varphi]}{\Delta_m-i\gamma/2},\ \ \ \ \ P^{\mathcal{C}}_{\rm T/R}=\sqrt{\frac{\gamma}{2}}\lim_{z\to 0}\sum_{m=1}^{N}\sum_{j=1}^{\infty}{\rm exp}[(-/+)im\varphi]\langle G|\sigma_mG^{(1)}_{\mathcal{C},j}(z)H_+|G\rangle.
	\end{align}
	The first term, $P^{\mathcal{I}}_{\rm T/R}$, involves only a sum over the qubit index. It represents the superposition of $N$ paths in which the $m$-th path corresponds to a photon emitted from the $m$-th qubit propagating along the waveguide without being reabsorbed by other qubits; we refer to this term as the probability amplitude of the ``non-interacting transition path". Consequently, the second term, $P^{\mathcal{C}}_{\rm T/R}$, represents the probability amplitude for the case in which the emitted photon can be reabsorbed by other qubits, and we refer to it as ``interacting transition path". Thus, the final probability amplitude for the single-photon scattering path, $\langle{\phi^1_{\pm}}|\psi^1\rangle$, is the combination of the non-interacting path and interacting paths. In Fig.\,\ref{fig2}, we present the single-photon scattering processes for an array with $N=3$ as an example.
	
	As for the two-photon scattering path, the conclusion is similar. In addition to the single-photon events, the two-photon scattering path, $\langle{\phi^2_{\pm}}|\psi^2\rangle$, further involves the two-excitation steady state of the qubit ensemble. After similar derivations, the probability amplitude for the two-photon scattering path can be expressed as
	\begin{align}
		\langle{\rm sup^2_{\pm}}|\psi^2\rangle=P^{\mathcal{I}\mathcal{I}}_{\rm T/R}+P^{\mathcal{I}\mathcal{C}}_{\rm T/R}+P^{\mathcal{C}\mathcal{I}}_{\rm T/R}+P^{\mathcal{C}\mathcal{C}}_{\rm T/R},
	\end{align}
	with
	\begin{equation}
		\left\{
		\begin{aligned}
			&P^{\mathcal{I}\mathcal{I}}_{\rm T/R}=\frac{\gamma^2}{2}\sum_{m>n}\frac{{\rm exp}[(0/2)i(m+n)\varphi]}{(\Delta_m-i\gamma/2)(\Delta_n-i\gamma/2)} \\
			&P^{\mathcal{I}\mathcal{C}}_{\rm T/R}=\gamma\lim_{z\to 0}\sum_{m>n}\sum_{j=1}^{\infty}{\rm exp}[(-/+)i(m+n)\varphi]\langle G|\sigma_m\sigma_nG^{(2)}_{\mathcal{I}}(z)H_+G^{(1)}_{\mathcal{C},j}(z)H_+|G\rangle \\
			&P^{\mathcal{C}\mathcal{I}}_{\rm T/R}=\gamma\lim_{z\to 0}\sum_{m>n}\sum_{j=1}^{\infty}{\rm exp}[(-/+)i(m+n)\varphi]\langle G|\sigma_m\sigma_nG^{(2)}_{\mathcal{C},j}(z)H_+G^{(1)}_{\mathcal{I}}(z)H_+|G\rangle \\
			&P^{\mathcal{C}\mathcal{C}}_{\rm T/R}=\gamma\lim_{z\to 0}\sum_{m>n}\sum_{j=1,k=1}^{\infty}{\rm exp}[(-/+)i(m+n)\varphi]\langle G|\sigma_m\sigma_nG^{(2)}_{\mathcal{C},j}(z)H_+G^{(1)}_{\mathcal{C},k}(z)H_+|G\rangle \\
		\end{aligned}
		\right..
	\end{equation}
	Similarly, the first term $P^{\mathcal{I}\mathcal{I}}_{\rm T/R}$ denotes the probability amplitude of the non-interacting path, while the remaining terms correspond to the probability amplitudes of the interacting path. As a consequence, NPPB in the transmission output involves nearly completely destructive interference among the single-photon scattering path with a probability amplitude $\langle{\phi^1_{+}}|\psi^1\rangle$, the two-photon scattering path with a probability amplitude $\langle{\phi^2_{+}}|\psi^2\rangle$, and the free propagation path with a probability amplitude $1$; whereas NPPB in the reflection output involves nearly completely destructive interference solely of the two-photon scattering path with a probability amplitude $\langle{\phi^2_{-}}|\psi^2\rangle$.
	
	In addition to the physical mechanism of NPPB, the mathematical structure of solutions satisfying $g_{\rm T/R}=0$ is somewhat subtle. For an array of $N$ qubits, the solutions for $g_{\rm T/R}=0$ form a $(N\!-\!2)$-dimensional submanifold. This is because, these solutions are essentially constrained by two conditions
	\begin{align}
		g_{\rm T/R}=0,\ \ \ \ \ \ \ \ \grad{g_{\rm T/R}}=\boldsymbol{0}.
	\end{align}
	The second condition arises from the fact that the correlation function is analytical and attains its minimum value at 0; hence, its gradient must vanish at 0. Consequently, solutions of $g_{\rm T/R}=\epsilon$ with $\epsilon\ne0$ form a $(N\!-\!1)$-dimensional submanifold, since only the condition $g_{\rm T/R}=\epsilon$ is imposed. However, as the value of $\epsilon$ decreases toward 0, one can expect that this $(N\!-\!1)$-dimensional submanifold nearly collapses into a $(N\!-\!2)$-dimensional submanifold, corresponding to the solutions of $g_{\rm T/R}=0$. As a result, solutions of $g_{\rm T}=10^{-10}$ for an array with $N=3$ form near a curve embedded in the 3D parameters space, as shown in the main text. In Fig.\,\ref{fig3}, we present solutions of the correlation functions for chains with $N\ge3$. The solutions are obtained as follows: (i) For $N=3$, we first use a nonlinear programming solver to find the maximum value of $\sum_{m}\Delta^2_m$, the set $\qty{\Delta_{1,\rm min},\ \Delta_{2,\rm min},\ \Delta_{3,\rm min}}$, under the constraint $g_{\rm T/R}=10^{-10}$. That is,
	\begin{align}
		g_{\rm T/R}\big|_{\Delta_1=\Delta_{1,\rm min},\Delta_2=\Delta_{2,\rm min},\Delta_3=\Delta_{3,\rm min}}=10^{-10},\ \ \ \ \ \ \sum_{m}\Delta^2_m\big|_{\Delta_1=\Delta_{1,\rm min},\Delta_2=\Delta_{2,\rm min},\Delta_3=\Delta_{3,\rm min}}=\min\qty{\sum_{m}\Delta^2_m}.
	\end{align}
	Based on this solution, we then apply the Gauss-Newton algorithm to automatically find solutions to the equation $g_{\rm T/R}=10^{-10}$ until a maximum number of solutions, $K_{\rm sols}$, is reached. Here, we set $K_{\rm sols}=10^5$. (ii) For $N>3$, we similarly find the minimum of $\sum_{m}\Delta^2_m$ under the constraints $g_{\rm T/R}=10^{-10}$; while in the Gauss-Newton algorithm, we fix the first $(N-3)$ detunings and solve the equation $g_{\rm T/R}=10^{-10}$ for the last three detunings. Due to the existence of NPPB, the PDF tends to be constant at $s\ll1$, with fluctuations near that constant, as shown in Figs.\,\ref{fig3}(a-f). These fluctuations arise from the standard deviation inherent in the numerical method used to estimate the PDF, and they can be effectively suppressed by increasing the number of disorder realizations [see Figs.\,\ref{fig3}(g-h)]. Details regarding the numerical method are discussed in the last section.

	\section{Correlation function in the weak- and strong-disorder limits}
	\subsection{Weak-disorder limit ($W\ll1$)}
	In the weak-disorder limit, one can expect that the photon correlations do not deviate significantly from their clean counterparts. This means that for the reflection output, the correlation functions are expected to be distributed around the value of $g_{\rm R}$ for $W=0$ [see Fig.\,\ref{fig4}(a)]; while for the transmission output, the correlation functions are expected to be distributed around $g_{\rm T}\sim+\infty$. To confirm this statement, we present the PDFs in Figs.\,\ref{fig4}(b-i). For the reflection output [see Figs.\,\ref{fig4}(f-i)], the shapes of $P(s)$ become increasingly sharper around $\qty{1,0.444,1,1}$, as disorder strength decreases. The peaks of the PDFs correspond to the values of the correlation function in the clean limit ($W=0$). For the transmission output, as the disorder strength decreases, when $\varphi=0.5\pi$ the region where $P(s)\ne0$ shifts further towards infinity; and when $\varphi=0$, the value of $P(s)$ decreases for smaller $s$ while it increases for larger $s$ [see Figs.\,\ref{fig4}(b-e)]. Both of these results indicate that $g_{\rm T}$ in the weak-disorder limit is distributed around infinity.
	\begin{figure*}
		\centering
		\includegraphics[width=18cm]{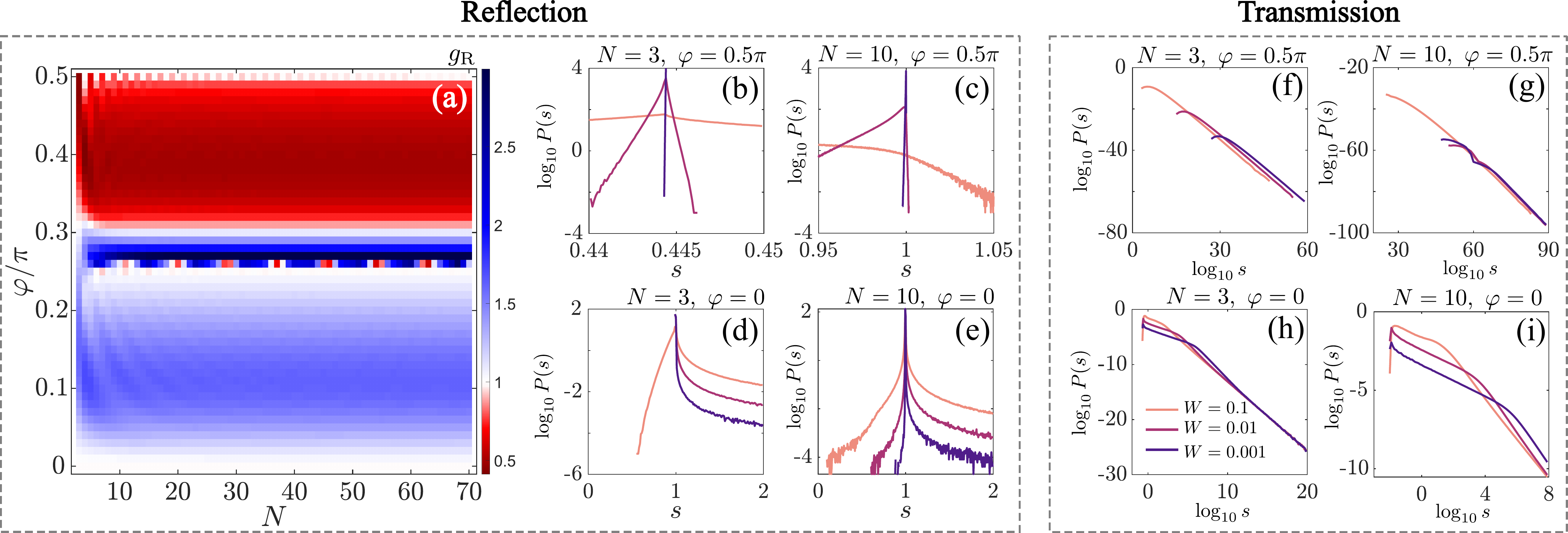}
		\caption{(a) Correlation functions for the reflection output in the absence of disorder. (b-e) PDFs for the reflection output versus disorder strength. Parameters are indicated above the plots. (f-i) PDFs for the reflection output versus disorder strength. Parameters are the same as (b-e). In all plots, the results are obtained from $10^{7}$ disorder realizations.}\label{fig4}
	\end{figure*}
	\begin{figure*}
		\centering
		\includegraphics[width=18cm]{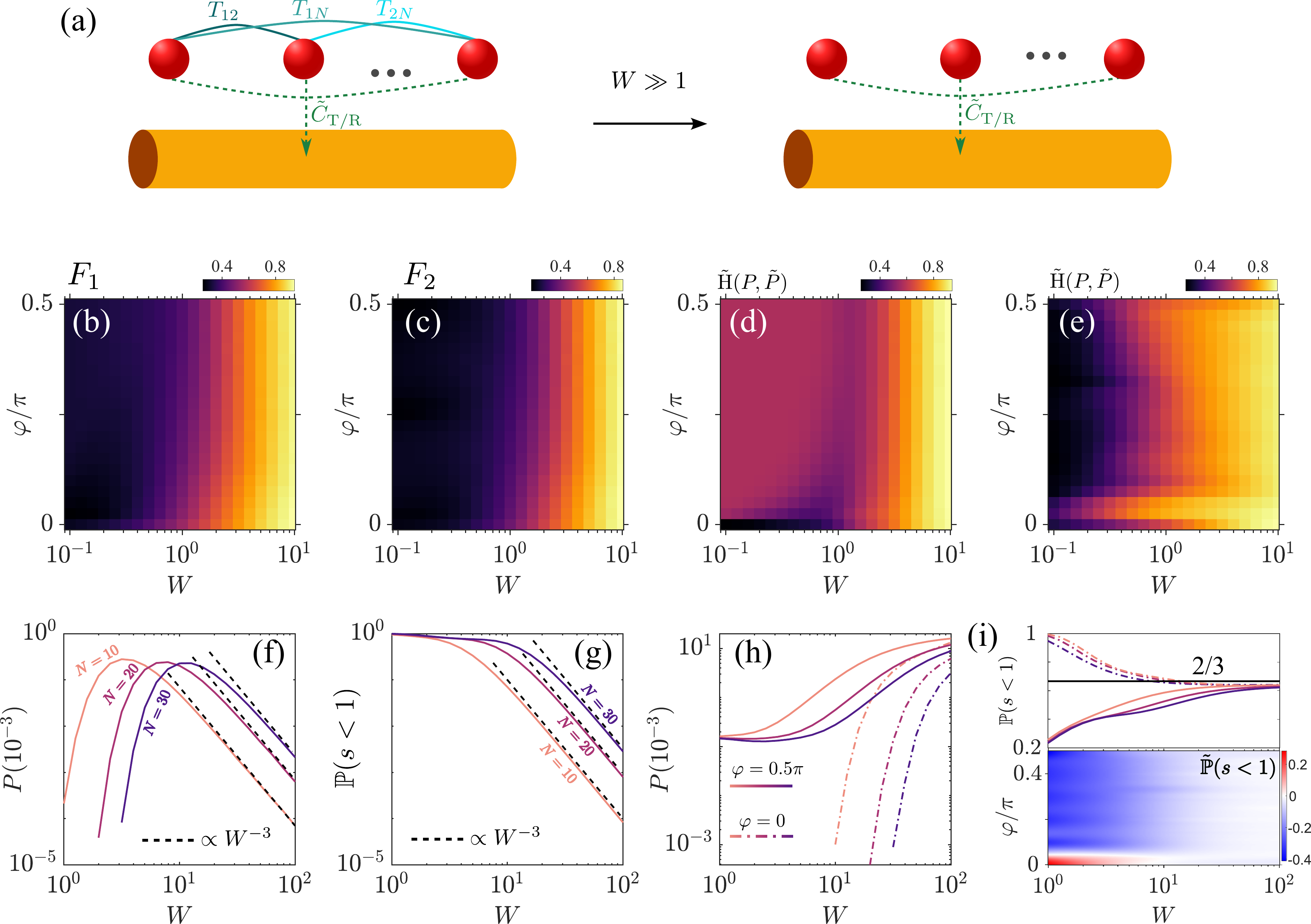}
		\caption{(a) Schematics of the systems in the weak- and intermediate-disorder regimes (left panel); and in the strong-disorder regime (right panel). Here $T_{mn}=-i \exp(i|m-n|\varphi)\sigma^{\dagger}_m\sigma_n/2$ with $m\ne n$. (b,c) Fidelity between the truncated steady state $\qty{|\psi^1\rangle,|\psi^2\rangle}$ obtained from the non-Hermitian Hamiltonian with long-range interactions and the truncated steady state $\qty{|\tilde{\psi}^1\rangle,|\tilde{\psi}^2\rangle}$ obtained from Eq.\,(\ref{psi_til}). (d,e) Hellinger distance between the PDFs obtained from Eqs.\,(\ref{eq9}-\ref{eq10}) and those obtained from Eqs.\,(\ref{eq30},\ref{eq32}). Here we plot $\tilde{\rm H}(P,\tilde{P})=1-\tilde{\rm H}(P,\tilde{P})$. $N=10$ in plots (b-e). The results are obtained from $10^{4}$ disorder realizations. (f,g) Probability of PA and $P(10^{-3})$ for the transmission output versus disorder strength for different system sizes. The dashed lines indicate slopes of $W^{-3}$. These results are obtained from Eq.\,(\ref{eq30}) using $10^{10}$ disorder realizations. (h,i) Probability of PA and $P(10^{-3})$ for the reflection output versus disorder strength for different system sizes and phases. These results are obtained from Eq.\,(\ref{eq32}) using $10^{10}$ disorder realizations. Different colored lines correspond to different chain sizes (the same as those in (f,g)). In the top panel of (i), the parameters are the same as in (h) (labeled within the plot), and in the bottom panel of (i), we plot $\mathbb{P}(s<1)-2/3$ for $N=10$.}\label{fig5}
	\end{figure*}
	\subsection{Strong-disorder limit ($W\gg1$)}
	In the strong-disorder limit, one can expect that the photon-mediated interaction between qubits is significantly quenched [see Fig.\,\ref{fig5}(a)]. This indicates that the probability amplitudes of interacting paths are extremely suppressed, so that one can consider only the probability amplitudes of non-interacting paths. In this situation, the correlation functions can be solved exactly, as derived in the previous section. After some calculations, the truncated steady states are given by
	\begin{align}\label{psi_til}
		|\tilde{\psi}^1\rangle=-\frac{\gamma}{2}\sum_{m=1}^{N}\frac{{\rm exp}(im\varphi)}{\Delta_m-i\gamma/2}\sigma^{\dagger}_m|G\rangle,\ \ \ \ \ \  |\tilde{\psi}^2\rangle=|\tilde{\psi^1}\rangle\otimes|\tilde{\psi^1}\rangle=\frac{\gamma^2}{2}\sum_{m>n}\frac{{\rm exp}(i(m+n)\varphi)}{(\Delta_m-i\gamma/2)(\Delta_n-i\gamma/2)}\sigma^{\dagger}_m\sigma^{\dagger}_n|G\rangle.
	\end{align}
	Thus,
	\begin{align}\label{eq30}
		g_{\rm T}=\frac{|1-2iP^{\mathcal{I}}_{\rm T}-P^{\mathcal{II}}_{\rm T}|^2}{|1-iP^{\mathcal{I}}_{\rm T}|^4}=\frac{|1+i\sum_m(\Delta_m-i/2)^{-1}-\sum_{m>n}(\Delta_m-i/2)^{-1}(\Delta_n-i/2)^{-1}/2|^2}{|1+i\sum_m(\Delta_m-i/2)^{-1}/2|^4}
	\end{align}
	and
	\begin{align}\label{eq32}
		g_{\rm R}=\frac{|P^{\mathcal{II}}_{\rm R}|^2}{|P^{\mathcal{I}}_{\rm R}|^4}=\frac{|2\sum_{m>n}\frac{e^{2i(m+n)\varphi}}{\qty(\Delta_m-i/2)\qty(\Delta_n-i/2)}|^2}{|\sum_m\frac{e^{2im\varphi}}{\Delta_m-i/2}|^4}.
	\end{align}
	From Eq.\,(\ref{eq30}), the correlation function in the transmission output is independent of the distance between qubits, i.e., independent of $\varphi$. In addition, solutions of $g_{\rm T}=0$ exist when $\varphi=0$, provided that $N\ge3$. For instance, one can easily verify that $g_{\rm T}=0$ if $\Delta_i=1/2$ and $\Delta_j=-1/2$ for $i\ne j$. However, these results do not contradict those obtained from Eq.\,(\ref{eq9}), where the correlation functions are distance-dependent and have no solutions satisfying $g_{\rm T}=0$ when $\varphi=0$. This discrepancy arises because Eqs.\,(\ref{eq30},\ref{eq32}) are valid only in the sense of disorder averaging.
	
	To examine the validity of Eqs.\,(\ref{eq30},\ref{eq32}), we numerically calculate the normalized fidelity
	\begin{align}
		F_1=\frac{|\langle \tilde{\psi}^1|\psi^1\rangle|}{\sqrt{|\langle \tilde{\psi}^1|\tilde{\psi}^1\rangle||\langle \psi^1|\psi^1\rangle|}},\ \ \ \ \ \ \ F_2=\frac{|\langle \tilde{\psi}^2|\psi^2\rangle|}{\sqrt{|\langle \tilde{\psi}^2|\tilde{\psi}^2\rangle||\langle \psi^2|\psi^2\rangle|}}.
	\end{align}
	Since photon correlations are fully encoded in the truncated steady states, the closer the normalized fidelity is to unity, the closer the values of the two types of correlation functions (obtained from Eqs.\,(\ref{eq9},\ref{eq10}) and Eqs.\,(\ref{eq30},\ref{eq32})) will be. We also investigate the Hellinger distance between the PDFs $P(s)$ obtained from Eqs.\,(\ref{eq9}-\ref{eq10}) and the PDFs $\tilde{P}(s)$ obtained from Eqs.\,(\ref{eq30},\ref{eq32}). The Hellinger distance between two probability density functions, $q(x)$ and $p(x)$, is defined as
	\begin{align}
		{\rm H}(q(x),p(x))=\frac{1}{2}\int\qty(\sqrt{p(x)}-\sqrt{q(x)})^2\dd{x}.
	\end{align}
	The Hellinger distance measures the similarity between $p(x)$ and $q(x)$; its values range from 0 to 1, with values closer to 0 indicating that the two distributions are more similar. From Figs.\,\ref{fig5}(b-e), one can see that (i) the fidelity is close to 1 for both the single- and two-excitation steady states as the disorder strength increases, and (ii) the value of ${\rm H}(P(s),\tilde{P}(s))$ approaches 0 both for both the transmission and reflection outputs as the disorder strength increases. This implies that Eqs.\,(\ref{eq30},\ref{eq32}) are indeed good approximations in the strong-disorder limit. Note that, although we only calculated the case of $N=10$ here, similar results can be obtained for other chain sizes.
	
	In Figs.\,\ref{fig5}(f-i), we present $P(10^{-3})$ and $\mathbb{P}(s<1)$ versus different disorder strengths. For the transmission output, our results indicate that both $P(10^{-3})$ and $\mathbb{P}(s<1)$ decrease as the disorder strength increases. This decrease follows a scaling of the form $W^{-3}$, regardless of the chain size. For the reflection output, the probability of NPPB increases with increasing disorder strength, while the probability of PA saturates to a constant in the limit $W\gg1$. This constant is close to $2/3$, i.e., the value of $\mathbb{P}(s<1)$ for a system with $N=2$ and $\varphi=0$. This result is also consistent with Fig.\,4(c) of the main text, where $\mathbb{P}(s<1)$ is calculated based on Eq.\,(\ref{gr_intef}).

	\begin{figure*}
		\centering
		\includegraphics[width=16cm]{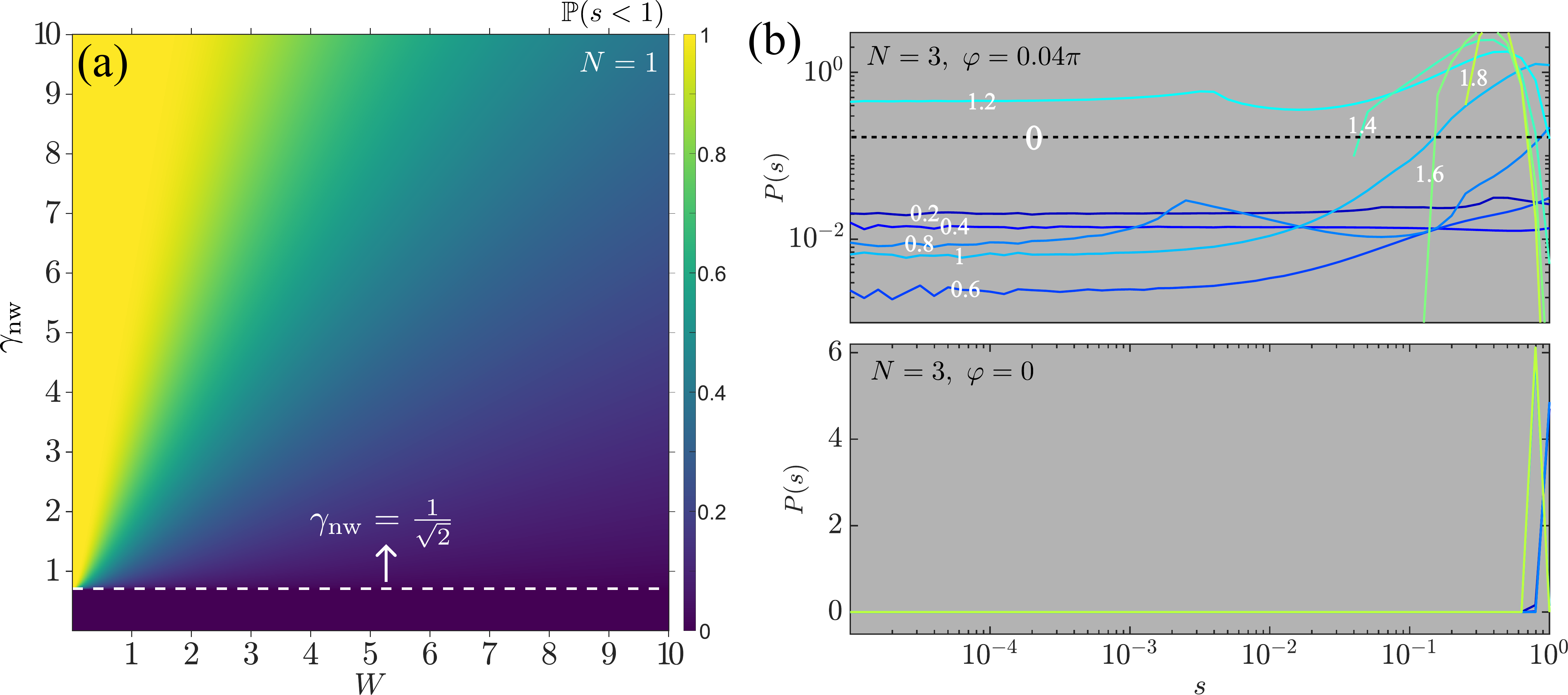}
		\caption{Correlation statistics of the transmission output. (a) Probability of PA versus $\gamma_{\rm nw}$ and $W$. Here $N=1$. (b) PDFs for different $\gamma_{\rm nw}$, with the values of $\gamma_{\rm nw}$ being indicated besides each curve. The chosen parameters are $\qty{N=3,\varphi=0.04\pi,W=0.14}$ (top) and $\qty{N=3,\varphi=0,W=1}$ (bottom), which are the same as Figs.\,2(a,b) of the main text. These results are obtained from $10^{10}$ disorder realizations.}\label{fig6}
	\end{figure*}
	
	\section{Effects of losses on non-waveguide modes}
	In this section, we consider losses to modes external to the waveguide. With such losses, the non-Hermitian Hamiltonian becomes
	\begin{align}
		H_{\rm eff}=\sum_{m,n=1}^{N}\qty(\Delta_m\delta_{m,n}-\frac{i\gamma_{\rm nw}}{2}\delta_{m,n}-\frac{i\gamma}{2}e^{i|m-n|\varphi})\sigma^{\dagger}_m\sigma_n,
	\end{align}
	where $\gamma_{\rm nw}$ denotes the decay rate to non-waveguide modes. Again, we set $\gamma=1$ in the remainder of this section and use the symbol $\beta$ to represent the coupling efficiency. The definition of $\beta$ is $\beta=\gamma/(\gamma+\gamma_{\rm nw})=(1+\gamma_{\rm nw})^{-1}$. In the main text, we set $\gamma_{\rm nw}=0$, corresponding to a unit value of the coupling efficiency.
	\subsection{Transmission output}
	We first consider the single-qubit system, for which $\mathbb{P}(s<1)$ and $P(0)$ can be analytically calculated. In this case, the correlation function is given by
	\begin{align}
		g_{\rm T}=\frac{(4\Delta^2_1+(\gamma_{\rm nw}-1)^2)(4\Delta^2_1+(\gamma_{\rm nw}+1)^2)}{(4\Delta^2_1+\gamma^2_{\rm nw})^2}.
	\end{align}
	Before proceeding to disordered systems, let us first focus on clean systems. For a resonant qubit, i.e., $\Delta_1=0$, the correlation function is $g_{\rm T}=(\gamma^2_{\rm nw}-1)^2/\gamma^4_{\rm nw}$, which satisfies $g_{\rm T}<1$ for $\gamma_{\rm nw}>1/\sqrt{2}$ ($\beta\lesssim 0.58$) and $g_{\rm T}=0$ for $\gamma_{\rm nw}=1$ ($\beta=1/2$). Unlike lossless system, here the transmission output can produce either antibunched or perfectly blockaded photons by appropriately adjusting the coupling efficiency.
	
	Considering the effects of disorder, the probability of PA is calculated as
	\begin{equation}
		\begin{aligned}
			\mathbb{P}(s<1)&=\frac{1}{\sqrt{2\pi}W}\int_{-\infty}^{\infty}e^{-\Delta^2_1/2W^2}\Theta\qty(\frac{(4\Delta^2_1+(\gamma_{\rm nw}-1)^2)(4\Delta^2_1+(\gamma_{\rm nw}+1)^2)}{(4\Delta^2_1+\gamma^2_{\rm nw})^2}-1)\dd{\Delta_1} \\
			&=\frac{1}{\sqrt{2\pi}W}\int_{0}^{\frac{\gamma^2_{\rm nw}}{4}-\frac{1}{8}}\frac{e^{-\Delta_1/2W^2}}{\sqrt{\Delta_1}}\Theta\qty(\gamma_{\rm nw}-\frac{1}{\sqrt{2}})\dd{\Delta_1} \\
			&=\erf\qty(\frac{\sqrt{\gamma^2_{\rm nw}/4-1/8}}{\sqrt{2}W})\Theta\qty(\gamma_{\rm nw}-\frac{1}{\sqrt{2}}),
		\end{aligned}
	\end{equation}
	where $\erf(x)$ is the error function. This result shows that when a single qubit is strongly coupled to the waveguide, PA in the transmission output is impossible. By reducing the coupling efficiency until reaching a critical value $\beta_{\rm cri}=1/(1+1/\sqrt{2})\approx 0.58$, PA events can be observed. For $\beta<\beta_{\rm cri}$, the probability of PA increases as the coupling efficiency decreases, as shown in Fig.\,\ref{fig6}(a).
	
	As for $P(0)$, we first calculate the PDF. After performing the integral (similar to Eq.\,(\ref{eq12})), the PDF for $s<1$ is given by
	\begin{equation}
		\begin{aligned}
			P(s)&=-\frac{1}{\sqrt{2\pi }W}\frac{(-1+\sqrt{s+4(1-s)\gamma^2_{\rm nw}})^3}{4\sqrt{-1+(s-1)\gamma^2_{\rm nw}+\sqrt{s+4(1-s)\gamma^2_{\rm nw}}}(s-1)^2(4(s-1)\gamma^2_{\rm nw}-s+\sqrt{s+4(1-s)\gamma^2_{\rm nw})}(1-s)^{3/2}}\times \\
			&\exp(\frac{-1+(s-1)\gamma^2_{\rm nw}+\sqrt{s+4(1-s)\gamma^2_{\rm nw}}}{(8s-8)W^2})
		\end{aligned},
	\end{equation}
	where $\gamma_{\rm nw}>1/\sqrt{2}$ and $1+\gamma^{-4}_{\rm nw}-2\gamma^{-2}_{\rm nw}<s<1$. From this result, $P(0)=0$ for $\beta\ne 1/2$ ($\gamma_{\rm nw}\ne1/2$); only when the coupling efficiency is exactly $1/2$, $P(s)$ diverges as $P(s)\sim (W\sqrt{s})^{-1}$, whereby PPB events become possible.
	
\begin{figure*}
		\centering
		\includegraphics[width=16cm]{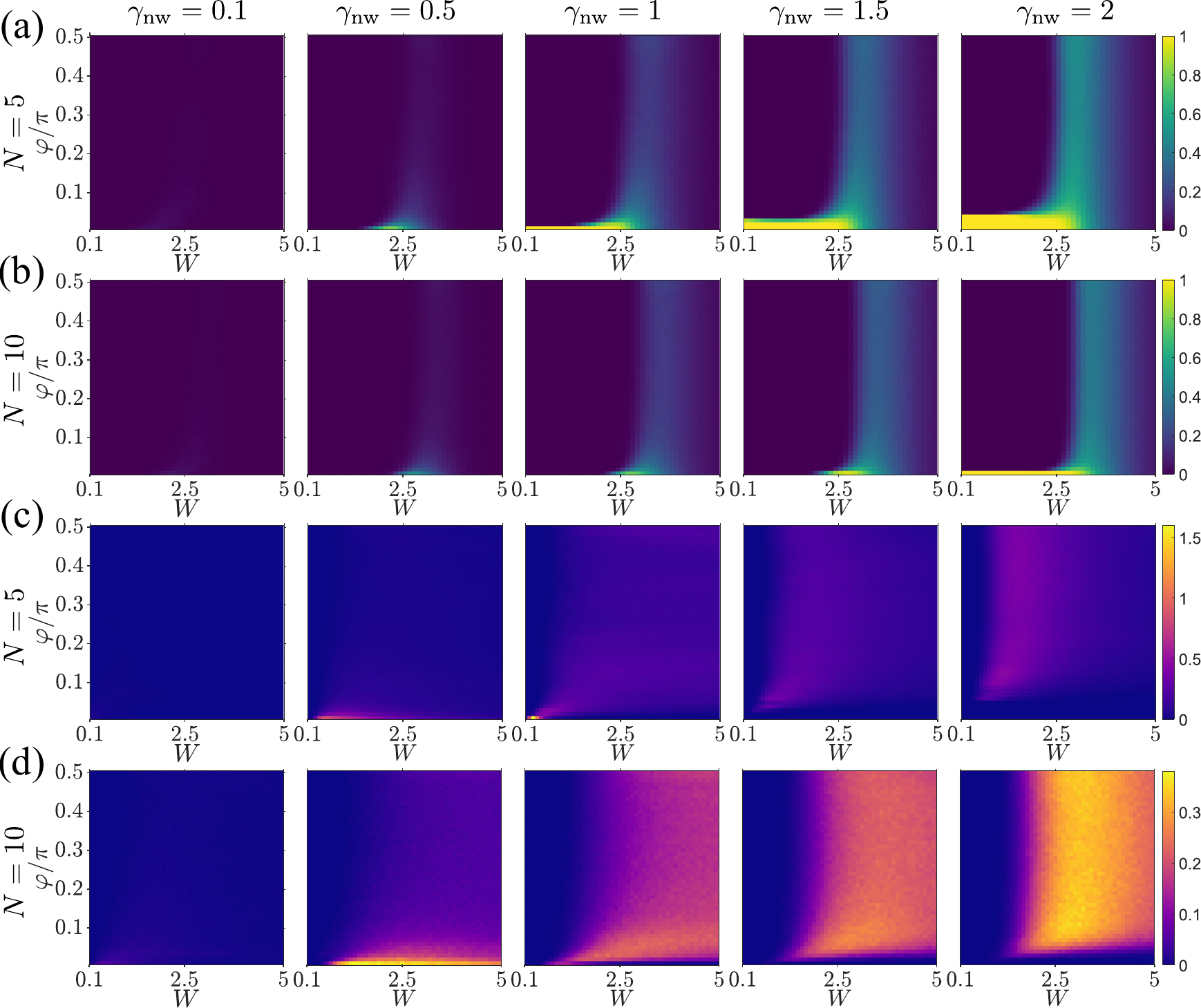}
		\caption{Correlation statistics of the transmission output. (a,b) Probability of PA versus $\varphi$ and $W$ for different $N$ and $\gamma_{\rm nw}$. (c,d) $P(10^{-3})$ versus $\varphi$ and $W$ for different $N$ and $\gamma_{\rm nw}$. These results for $\mathbb{P}(s<1)$ ($P(10^{-3})$) are obtained from $5\time 10^{4}$ ($5\times10^{6}$) disorder realizations.}\label{fig7}
	\end{figure*}

	In Figs.\,\ref{fig7}(a,b), we present the numerical results for the probability of PA for $N=5$ and $N=10$. These results show that the probability of PA can be efficiently enhanced by decreasing the coupling efficiency (by increasing $\gamma_{\rm nw}$). In particular, the probability of PA can reach unit provided that the chain is highly dense, i.e., $\varphi\ll1$. We then calculate the PDFs for the system with $N=3$ [see Fig.\,\ref{fig6}](b). When $\varphi\ne0$, the results show that as the coupling efficiency decreases, the PDFs still exhibit a constant behavior at $s\ll1$ as long as $\beta\gtrsim 0.45$; this behavior disappears for $\beta\lesssim0.45$. For $\beta\gtrsim0.45$, the value of $P(s)$ at $s\ll1$ can even be larger than that for lossless systems ($\gamma_{\rm nw}=0$). This is observed, for instance, when $\gamma_{\rm nw}=1.2$ ($\beta\approxeq 0.45$). Therefore, we still use $P(10^{-3})$ as a measure for the probability of NPPB and present the results for different system parameters in Figs.\,\ref{fig7}(c,d). The results show that the probability of NPPB attains a larger value when the coupling efficiency is not too high ($\beta\lesssim0.5$). Moreover, when $\varphi=0$, the generation of strongly antibunched photon is impossible [see Fig.\,\ref{fig6}], similar to the system with $\beta=1$.
	
	\subsection{Reflection}
\begin{figure*}
		\centering
		\includegraphics[width=16cm]{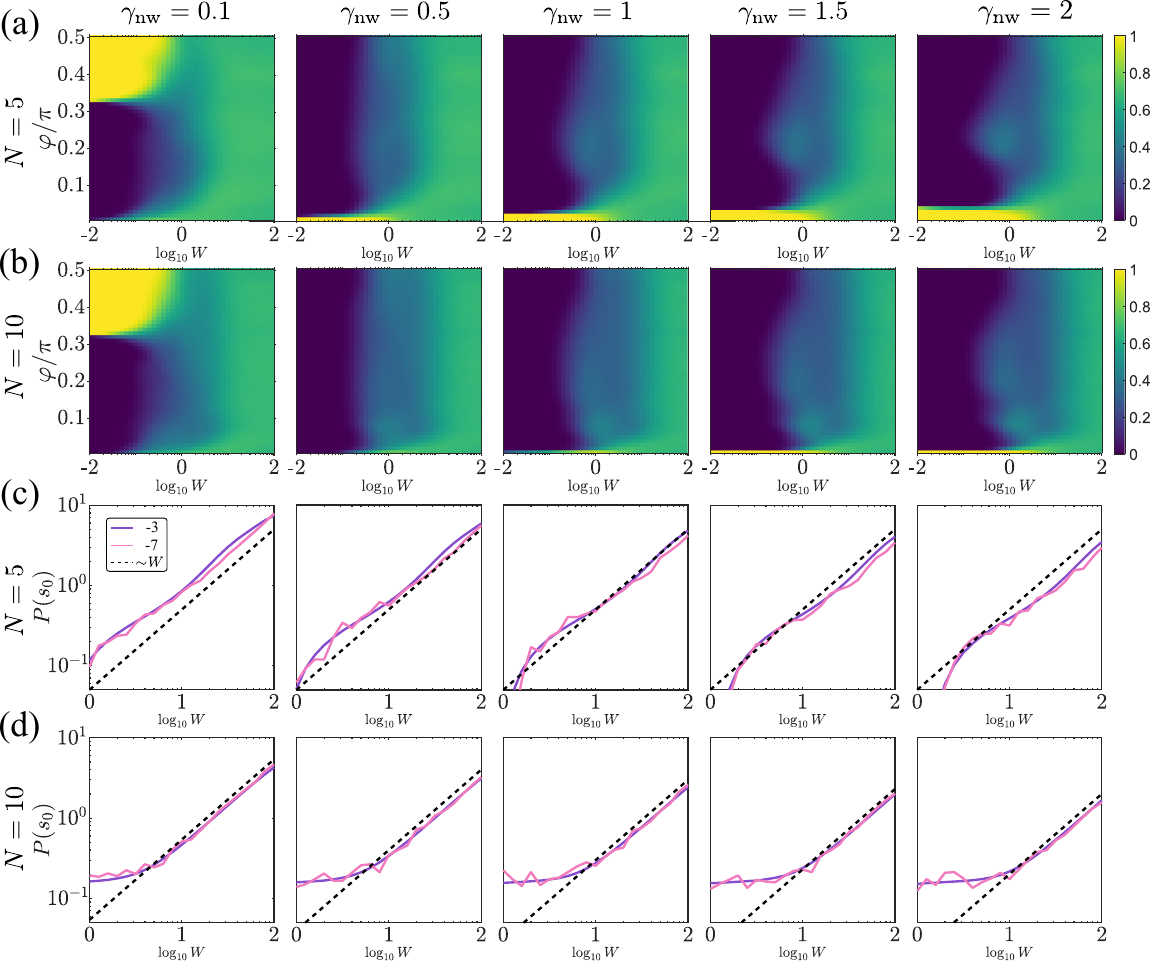}
		\caption{Correlation statistics of the reflection output. (a,b) Probability of versus $\varphi$ and $W$ for different $N$ and $\gamma_{\rm nw}$. (c,d) Solid lines denote $P(10^{\mu})$ versus $\varphi$ and $W$ for different $N$ and $\gamma_{\rm nw}$, with the values of $\mu=-3,\ -7$. Dashed lines represent the numerical fits of $\sim W$. These results for $\mathbb{P}(s<1)$ ($P(s_0)$) are obtained from $5\time 10^{4}$ ($5\times10^{10}$) disorder realizations.}\label{fig8}
	\end{figure*}

	In Figs.\,\ref{fig8}(a,b), we present the results of the probability of PA for $N=5$ and $N=10$. As the coupling efficiency decreases, a high probability of PA can only be achieved when the chain is highly dense, which is similar to the transmission output. Besides, it maintains a finite value even for strong disorder strength. This is because the effects of losses will be suppressed by strong disorder. We also investigate the behavior of $P(s_0)$ at $s_0\ll1$ when $W\gg1$. In this limit, the correlation function is given by
	\begin{align}\label{eqbeta}
		g_{\rm R}=\frac{|P^{\mathcal{II}}_{\rm R}|^2}{|P^{\mathcal{I}}_{\rm R}|^4}=\frac{|2\sum_{m>n}\frac{e^{2i(m+n)\varphi}}{\qty(\Delta_m-i/2-i\gamma_{\rm nw}/2)\qty(\Delta_n-i/2-i\gamma_{\rm nw}/2)}|^2}{|\sum_m\frac{e^{2im\varphi}}{\Delta_m-i/2-i\gamma_{\rm nw}/2}|^4}.
	\end{align}
	The numerical results in Figs.\,\ref{fig8}(c,d) show clearly that the probability of NPPB increases with increasing disorder strength, which is similar to systems where $\beta=1$.
	
	
	\section{Effects of chirality in coupling to waveguide modes}
\begin{figure*}
		\centering
		\includegraphics[width=16cm]{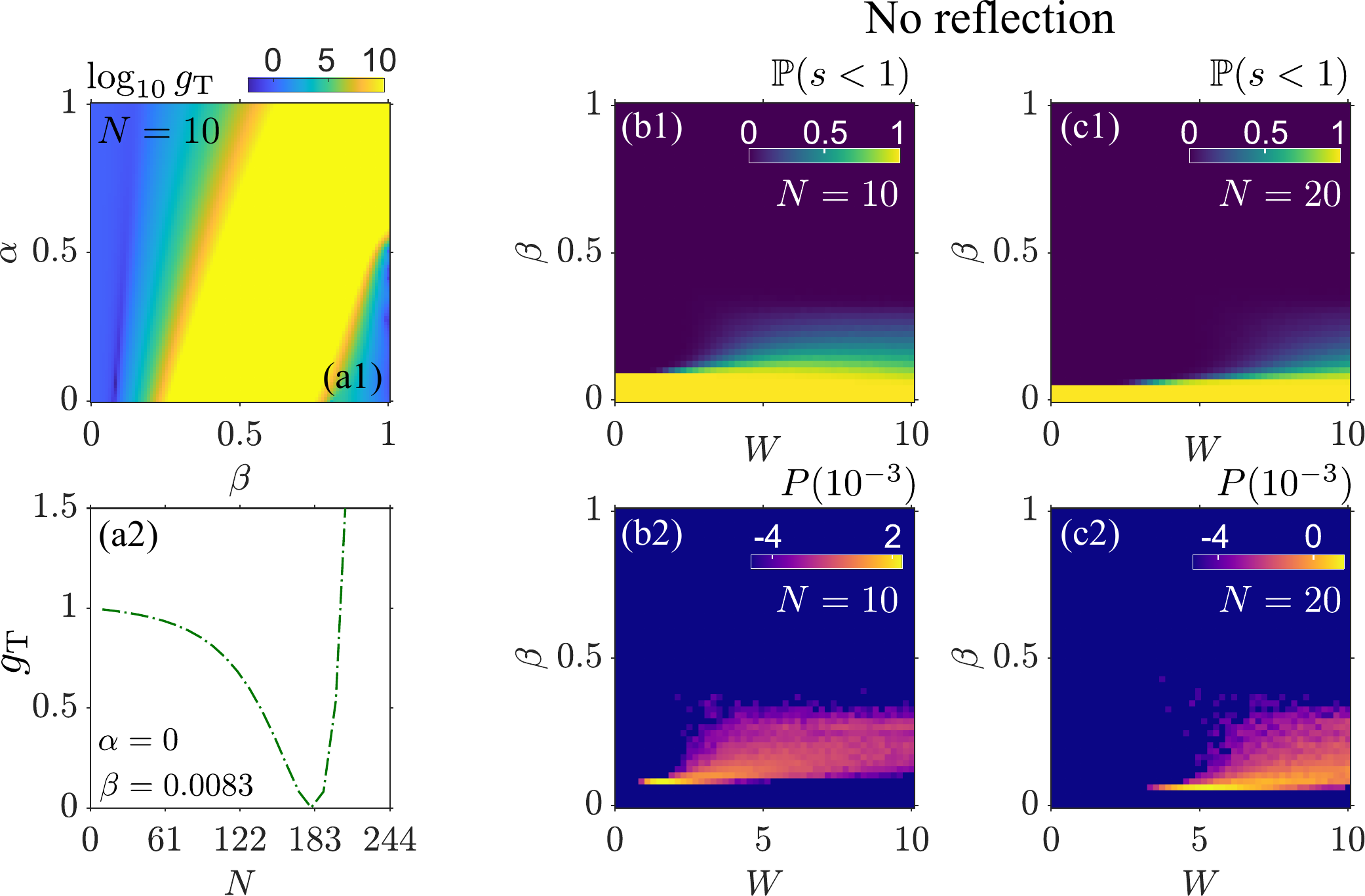}
		\caption{(a1) Photon correlations in the transmission output versus chirality and coupling strength for $N=10$ and $\varphi=0$. (a2) Photon correlations in the transmission output versus chain size. Here the chosen parameters are $\alpha=0$ and $\beta=0.0083$. (b1,c1) Probability of photon antibunching in the transmission output versus disorder and coupling strength. $N=10$ in (b1) and $N=20$ in (c1). (b2,c2) $P(10^{-3})$ in the transmission output versus disorder and coupling strength. $N=10$ in (b2) and $N=20$ in (c2). $\alpha=0$ in (b1-c2). Results are obtained from $10^5$ disorder realizations.}\label{fig11}
	\end{figure*}

	In this section, we consider chirality in coupling to waveguide modes, i.e., $\gamma_{\rm T}\ne\gamma_{\rm R}$. We define the chirality as $\alpha=\gamma_{\rm R}/\gamma_{\rm T}$, and we set $\gamma_{\rm T}=1$ as the energy unit. Figures\,\ref{fig11}(a1) show the photon correaltions for transmission output, for system with chain size $N=10$. The chirality together with the losses on non-waveguide modes significantly affect photon correlations. For transmission output, when qubits weakly coupled to waveguide ($\beta\ll1$), photons maintain the coherence for all values of chirality; while when qubits strongly coupled to waveguide ($\beta\approx 1$), photons exhibit strong bunching for $\alpha\gtrsim 0.5$, and antibunched even strongly antibunched photon emerge by further increasing chirality (decreasing $\alpha$). When the qubits are weakly coupled to the waveguide in a perfectly chiral fashion, the transmitted photon exhibit Poisson statistics for small system size, $N=10$. As the chain size increases, the output becomes antibunched and approaches near-perfect antibunching at the optimal chain size $N\approx 183$; beyond this optimal chain size the output light becomes bunched as $N$ increases further [see Fig.\,\ref{fig11}(a2)]. Note that these behaviors are quantitatively consistent with Fig.\,3a in\,\cite{prasad_correlating_2020_supp}.
	
	
	In the presence of disorder, Figs.\,\ref{fig11}(b1-c2) show $\mathbb{P}(s<1)$ and $P(10^{-3})$ for the case of complete transimission-chirality ($\alpha=0$). On the one hand, the probability of photon antibunching approach unit when qubits weakly coupled to waveguide, and rapidly decrease with increasing coupling strength. On the other hand, the probability of strong photon antibunching exhibits non-negligible value only for $0.1\lesssim\beta\lesssim 0.4$. Notably, for the chain size investigated here $N=10,20$, these behaviors of $\mathbb{P}(s<1)$ and $P(10^{-3})$ do not change significantly with changing $N$. Considering that when $N\sim 10^2$, the qubit number significantly affects the correlations even without the presence of disorder, we expect that $\mathbb{P}(s<1)$ and $P(10^{-3})$ may show quite distinctive behaviors with further increasing $N$, in comparison with the results provided here.
	
	\section{Effects of finite bandwidth of input state}
	In this section, we consider the effects of finite bandwidth of the input state, which should be compared with the single-mode (zero bandwidth) coherent input state considered in the main text. We assume that the input state has the form
	\begin{align}
		\rho_E(t_0)\propto \exp\qty(\bar{n}\int_{-\infty}^{\infty}\alpha(\omega)a_{\rm T}(\omega)-\alpha(\omega)a^{\dagger}_{\rm T}(\omega)\dd{\omega}).
	\end{align}
	Here, $\alpha(\omega)$ controls the profile of the input state and is required to be normalized, i.e., $\int (\alpha(\omega))^2\dd{\omega}=1$, such that $\bar{n}^2=\int_{-\infty}^{\infty}\Tr\qty[\rho_E a^{\dagger}_{\rm T}(\omega)a_{\rm T}(\omega)]\dd{\omega}$ represents the total number of photons in the input state, and we still consider the weak-input limit $\bar{n}^2\ll1$. To be able to compare with the case of zero bandwidth, we require $\alpha(\omega)$ to exhibit two following generic features: (i) $\alpha(\omega)$ approaches its maximum at $\omega_0$ and rapidly decreases to zero for $\omega$ away from $\omega_0$; (ii) a characteristic bandwidth denoted by $\sigma_{\omega}$ is able to characterize the width of $\alpha(\omega)$, such that $\sigma_{\omega}\to 0$ ($\sigma_{\omega}\to \infty$) corresponds to a ideal zero-bandwidth input ($\delta$-pulse input in the time domain). Hereafter, we consider $\alpha(\omega)$ to be Lorentz profile, i.e., (we assume $\gamma=1$)
	\begin{align}
		\alpha(\omega)=\frac{1}{\mathcal{N}}\frac{\sigma_{\omega}}{(\omega-\omega_0)^2+\sigma^2_{\omega}},
	\end{align}
	where $\mathcal{N}=(\pi/2\sigma_{\omega})^{1/2}$ denotes the normalization factor. The total number of photons in the time domain is given by
	\begin{align}
		\bar{n}^2(t)=\int\Tr{a^{\dagger}_{\rm T}(\omega,t)a_{\rm T}(\omega,t)\rho_E(t_0)}\dd{\omega}=\frac{1}{\sqrt{2\pi}}\int\Tr{e^{-i\omega (t-t_0)}a^{\dagger}_{\rm T}(\omega)a_{\rm T}(\omega)\rho_E(t_0)}\dd{\omega}=\bar{n}^2\frac{e^{-\sigma_{\omega}\abs{t-t_0}}(1+\sigma_{\omega}\abs{t-t_0})}{\sqrt{2\pi}}.
	\end{align}
	This expression demonstrates that the input is also a wave package in the time domain, where $\bar{n}^2(t)$ approaches its maximum at $t=t_0$ and rapidly decreases for $t$ away from $t_0$.
	
	In this case, the master equation is still given by Eq.\,(\ref{ME}), with
	\begin{align}
		f(t-t_0,md)=\bar{n}\sqrt{\sigma_{\omega}}e^{-\sigma_{\omega}\abs{t-t_0-md/v_g}}e^{-i\omega_0(t-t_0-md/v_g)}\approx \bar{n}\sqrt{\sigma_{\omega}}e^{-\sigma_{\omega}\abs{t-t_0}}e^{-i\omega_0(t-t_0)},
	\end{align}
	where we assume that photons are injected far from the atomic ensemble, such that $t_0+ md/v_g\approx t_0$. Consequently, the correlation function can be obtained as
	\begin{align}\label{cor_sigma}
		g_{\mu}(\tau,\tau)=\frac{\Tr\qty[\rho(\tau)a^{\dagger}_{\mu,\rm out}(\tau)a^{\dagger}_{\mu,\rm out}(\tau)a_{\mu,\rm out}(\tau)a_{\mu,\rm out}(\tau)]}{\Tr\qty[\rho(\tau)a^{\dagger}_{\mu,\rm out}(\tau)a_{\mu,\rm out}(\tau)]^2},
	\end{align}
	where $\rho(\tau)$ denotes the density matrix of atomic ensemble governed by the master equation. The input-output relations read as
	\begin{align}
		a_{\rm T, out}(\tau)=a_{\rm T, in}(\tau)-i\sqrt{\frac{\gamma}{2}}\sum_me^{-imd}\sigma_m(\tau),\ \ \ \ a_{\rm R, out}(\tau)=-i\sqrt{\frac{\gamma}{2}}\sum_me^{imd}\sigma_m(\tau),
	\end{align}
	with
	\begin{align}
		a_{\rm T,in}(\tau)=\frac{1}{\sqrt{2\pi}}\int\Tr\qty[\rho_E(t_0)e^{-i\omega(\tau-t_0)}a_{\rm T}(\omega,t_0)]\dd{\omega}=\bar{n}\sqrt{\sigma_{\omega}}e^{-\sigma_{\omega}\abs{\tau-t_0}}e^{-i\omega_0(\tau-t_0)}.
	\end{align}
	
	Before proceeding to present the results for the correlation functions, we stress that, in addition to the bandwidth $\sigma_{\omega}$, the evolution time $\tau$ will also significantly affect the photon correlation. For a ideal zero-bandwidth input, the correlation functions $g_{\mu}$ are time-independent for $t\gg1$, and thus their values can be obtained from the steady state of the atomic ensemble. This actually arises from the fact that the drive strength $\propto|f(t-t_0,md)|$ remains a constant in the time domain. However, for a finite-bandwidth input, the drive strength is now time-dependent, therefore, one should expect that $g_{\mu}(\tau,\tau)$ is also dependent on the evolution time $\tau$.
	
\begin{figure*}
		\centering
		\includegraphics[width=16cm]{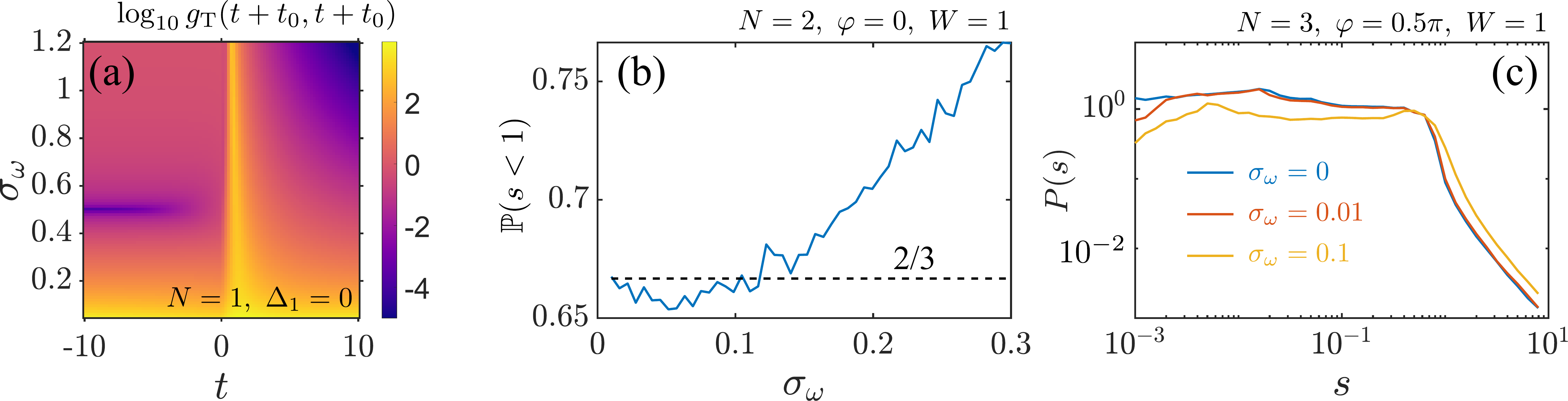}
		\caption{(a) Photon correlation in the transmission versus bandwidth and evolution time for system with $N=1$ and $\Delta_1=0$. (b) Probability of photon antibunching in the reflection output versus band width for system with $N=2$ and $\varphi=0$. Results are obtained from $10000$ disorder realizations. (c) Probability density function in the reflection output for different values of band width for system with $N=3$ and $\varphi=0.5\pi$. Results are obtained from $10^5$ disorder realizations. $\tau=t_0$ in (b) and (c). $\bar{n}=0.1$ in (a-c).}\label{fig10}
	\end{figure*}

	In Fig.\,\ref{fig10}(a), we present the photon correlation in the transmission output for $N=1$ with $\Delta_1=0$. For a ideal zero-bandwidth input, $g_{T}=\infty$. Our result reveals that $g_{\rm T}(\tau,\tau)$ generally recovers its zero-bandwidth counterpart when $\sigma_{\omega}\ll1$. This is because, when the bandwidth is much smaller than the individual decay rate $\gamma$, the input can be approximately considered as a zero-bandwidth coherent state, resulting in $g_{\rm T}(\tau,\tau)\gg1$. Increasing $\sigma_{\omega}$, $g_{\rm T}(\tau,\tau)$ becomes sensitive to the evolution time $\tau$: its value still approximately recovers the zero-bandwidth counterpart for $\tau\sim t_0$; while for $|\tau-t_0|\gg1$, $g_{\rm T}(\tau,\tau)$ significantly deviates from infinity, which can even approaches near zero. In Figs.\,\ref{fig10}(b,c), we present the probability of PA for system with $N=2$ and $\varphi=0$, and the probability density functions for system with $N=3$ and $\varphi=0.5\pi$ in the reflection output. The obtained results show good agreement with their zero-bandwidth counterpart as long as $\sigma_{\omega}/\gamma\lesssim 0.1$, i.e., $\mathbb{P}(s<1)=2/3$ and $P(s\ll1)\ne 0$.
	
	In summary, when one considers the effect of finite band width of input state, the obtained results show good agreement with their zero-bandwidth counterpart, as long as the bandwidth $\sigma_{\omega}$ is much smaller than the individual decay rate $\gamma$, and the detection time $\tau$ is appropriately chosen. In typical waveguide QED platforms, $\gamma\sim {\rm MHz}$, therefore, the required band width should be $\sigma_{\omega}\lesssim {\rm KHz}$, which can be achieved in the state-of-art waveguide platforms\,\cite{faez2014coherent_supp,kuyken2015octave_supp,bao2024cryogenic_supp}.
	
	\section{Numerical calculations of $\mathbb{P}(s<1)$ and $P(s\ll1)$ for many-qubit system}
	\subsection{Calculation about $\mathbb{P}(s<1)$}
\begin{figure*}
		\centering
		\includegraphics[width=16cm]{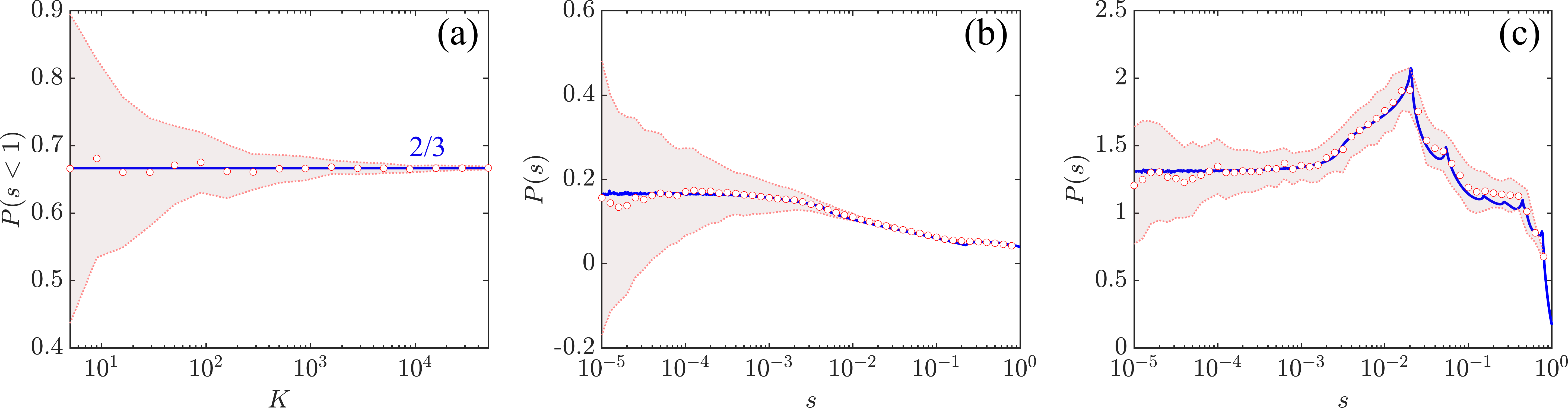}
		\caption{(a) Probability of PA for the reflection output from the Monte Carlo integration. The result is obtained from $10^2$ estimations. Red dots and the shaded area denote the mean value and the standard deviation of the estimations, respectively. Here the chosen parameter are $N=2$ and $\varphi=0$, so that the solid line represents the exact value of $\mathbb{P}(s<1)$, which is equal to $2/3$. (b-c) PDFs for the transmission (b) and reflection (c) outputs from the Monte Carlo integration. The result is obtained from $10^2$ estimations. Red dots and the shaded area denote the mean value and the standard deviation from Eq.\,(\ref{eq46}), respectively. The number of disorder realizations is $K=50000$. The solid blue lines represent the results from $K=10^{11}$ disorder realizations. Here the chosen parameters are $N=3$, $\varphi=0.04\pi$, and $W=0.15$ in (b); $N=3$, $\varphi=0.5\pi$, $W=1$ in (c).}\label{fig9}
	\end{figure*}

	The definition of $\mathbb{P}(s<1)$ is
	\begin{align}
		\mathbb{P}(s<1)=\int_0^1 P(s){\rm d}s =\int_{-\infty}^{\infty}\!\cdots\!\int_{-\infty}^{\infty} \Theta\qty(g_{\mu}-1)p(\Delta_1,\Delta_2,\!\cdots\!,\Delta_N) {\rm d}\Delta_1\cdots{\rm d}\Delta_N.
	\end{align}
	According to the Monte Carlo method, this integral can be approximately evaluated by
	\begin{align}\label{eq45}
		\mathbb{P}(s<1)= E\qty[\Theta\qty(g_{\mu}-1)]\approx \frac{1}{K}\sum_{j=1}^K\Theta\qty(g_{\mu}\big|_{\vec{\Delta}_j}-1),
	\end{align}
	where $\vec{\Delta}_j=\qty{\Delta_{1,j},\cdots,\Delta_{N,j}}$ denotes the $j$-th sample drawn from the i.i.d Gaussian distribution. The sample size $K$ should be large enough to reduce the variance of the estimation in Eq.\,(\ref{eq45}), which is given by $V[\Theta\qty(g_{\mu}-1)]\sim K^{-1}$. In Fig.\,\ref{fig9}(a), we present $E\qty[\Theta\qty(g_{\mu}-1)]$ for the reflection output versus the sample size $K$, with $N=2$ and $\varphi=0$. The mean and standard deviation are obtained from $10^2$ estimations of Eq.\,(\ref{eq45}). As shown, $K\sim 10^4$ is sufficient to reduce $V[\Theta\qty(g_{\mu}-1)]$, so that the mean value deviate only slightly from the exact value of $2/3$, even for a single estimation. Therefore, in the main text, we set $K=50000$ and perform a single estimation for all plots concerning $\mathbb{P}(s<1)$.
	
	\subsection{Calculation about $P(s\ll1)$}
	Similarly, $P(s)$ can be approximately estimated by the Monte Carlo method according to
	\begin{align}\label{eq46}
		P(s)=E[\delta\qty(g_{\mu}-s)]\approx \frac{1}{K}\sum_{j=1}^K\lim_{\varepsilon\to 0} \tilde{\delta}_{\varepsilon}\qty(g_{\mu}\big|_{\vec{\Delta}_j}-s),
	\end{align}
	where $\tilde{\delta}_{\varepsilon}(x)$ represents a function that weakly converges to the Dirac-delta function in the limit $\varepsilon\to0$. Compare to the estimation in Eq.\,(\ref{eq45}), the estimation of $P(s)$, especially at $s\ll1$, requires a larger number of disorder realizations to achieve an acceptable error. This is because the number of events $s-\epsilon<g_{\mu}<s+\epsilon$ with $\epsilon\ll s$ decreases with decreasing $s$. To see this, note that the probability of the event $s_0-\epsilon<g_{\mu}<s_0+\epsilon$, with $\epsilon\ll s_0$, is given by
	\begin{align}
		\mathbb{P}(s_0-\epsilon<s<s_0+\epsilon)=\int_{s_0-\epsilon}^{s_0+\epsilon} P(s)\dd{s}\sim P(s_0)\epsilon.
	\end{align}
	Since we have shown that $P(s_0)\sim{\rm constant}$ for $s_0\ll1$, this probability decreases as $s_0$ decreases, due to the condition $\epsilon\ll s_0$. In fact, the constant form of $P(s)$ implies that this probability scales as $s$; i.e., $\mathbb{P}(s_0-\epsilon<s<s_0+\epsilon)\sim s$ if we let $\epsilon=\eta s_0$ with $\eta\ll1$. The decrease of $\mathbb{P}(s_0-\epsilon<s<s_0+\epsilon)\sim s$ means that the number of events with $s-\epsilon<g_{\mu}<s+\epsilon$ also decreases as $s$ decreases. As a result, many estimations from Eq.\,(\ref{eq46}) for $s\ll1$ will yield zero if $K$ is not sufficiently large, because the value of $\tilde{\delta}_{\epsilon}(x)$ is proportional to the number of such events. Meanwhile, the computational cost increases rapidly with $K$, especially for large chain size. To balance the accuracy of the Monte Carlo integration results with the computational cost, an appropriate value of $K$ must be chosen.
	
	In Figs.\,\ref{fig9}(b,c), we compare the PDFs obtained from $K=50000$ with those obtained from $K=10^{11}$. For $K=50000$, we perform $10^{2}$ estimations so that the total sample size is $5\time10^{6}$. For $K=10^{11}$, the variation in the estimation is negligible, and it can be regarded as the exact PDF. Compared to the estimation of Eq.\,(\ref{eq45}), the variation increases as $s$ decreases $s$ and becomes non-negligible for $s\ll1$. However, despite the increased variation, the mean values from $10^{2}$ are very close to the exact value. Therefore, unless otherwise noted, we set $K=50000$ and perform $10^2$ estimations using Eq.\,(\ref{eq46}) to obtain all plots related to $P(10^{-3})$ in the main text.

	%

\end{document}